\documentclass[aps,prd,twocolumn,showpacs,superscriptaddress,groupedaddress]{revtex4-1}

\pdfoutput=1

\usepackage{graphicx}  
\usepackage{dcolumn}   
\usepackage{bm}        
\usepackage{amssymb}   
\usepackage{amsmath}   

\usepackage{color} 

\usepackage{multirow}
\usepackage{enumerate}
\usepackage{comment}
\usepackage{tabularx}
\usepackage{booktabs}

\usepackage{gensymb}
\usepackage{siunitx}

\def\dd{\operatorname{d}}
\def\msolar{\operatorname{M\textsubscript{\(\odot\)}}}
\def\Mpc{\operatorname{Mpc}}

\def\Mtot{\operatorname{\textit{M}_\text{tot}}}
\def\Mc{\operatorname{\textit{M}_\text{c}}}
\def\tc{\operatorname{\textit{t}_\text{c}}}

\begin{document}

\title{Multiband Gravitational Wave Parameter Estimation: A Study of Future Detectors}

\author{Stefan Grimm}
\email{stefan.grimm@gssi.it}
\affiliation{Gran Sasso Science Institute (GSSI), I-67100 L'Aquila, Italy}
\affiliation{INFN, Laboratori Nazionali del Gran Sasso, I-67100 Assergi, Italy}
\author{Jan Harms}%
\email{jan.harms@gssi.it}
\affiliation{Gran Sasso Science Institute (GSSI), I-67100 L'Aquila, Italy}
\affiliation{INFN, Laboratori Nazionali del Gran Sasso, I-67100 Assergi, Italy}

\date{\today}

\begin{abstract}
The first detection of a gravitational-wave signal of a coalescence of two black holes marked the beginning of the era of gravitational-wave astronomy, which opens exciting new possibilities in the fields of astronomy, astrophysics and cosmology. The currently operating detectors of the LIGO and Virgo collaborations are sensitive at relatively high frequencies, from 10\,Hz up to about a kHz, and are able to detect gravitational waves emitted in a short time frame of less than a second (binary black holes) to minutes (binary neutron stars). Future missions like LISA will be sensitive in lower frequency ranges, which will make it possible to detect gravitational waves emitted long before these binaries merge. In this article, we investigate the possibilities for parameter estimation using the Fisher-matrix formalism with combined information from present and future detectors in different frequency bands. The detectors we consider are the LIGO/Virgo detectors, the Einstein Telescope (ET), the Laser Interferometer Space Antenna (LISA), and the first stage of the Deci-Hertz Interferometer Gravitational wave Observatory (B-DECIGO). The underlying models are constructed in time domain, which allows us to accurately model long-duration signal observations with multiband (or broadband) detector networks on parameter estimation. We assess the benefit of combining information from ground-based and space-borne detectors, and how choices of the orbit of B-DECIGO influence  parameter estimates.
\end{abstract}

\maketitle

\section{Introduction}

The detection of binary black-hole (BBH) mergers in the LIGO/Virgo detectors and the reconstruction of their properties has opened up exciting new possibilities in astrophysics \cite{BBHdetection1,BBHdetection2,BBHdetection3,BBHdetection4}. Gravitational waves are a new observational window, which will allow us to gain more information about the properties of astrophysical objects like black holes or neutron stars. The observation of the mergers of compact objects is a unique testbed for general relativity in the strong-field regime, which will help to constrain alternative theories of gravity \cite{YYP2016, AbEA2016c, Wil1998}. The information we can obtain from binary mergers in the LIGO/Virgo detectors is, however, limited in many respects. With respect to the determination of the mass parameters, large uncertainties remain, due to the fact that the signal is only observed for a few seconds. While it is possible to determine the sky localization of the source with a three-detector network, multi-messenger observations are hindered by the fact that the sky localization of the source can be determined only \textit{after} the merger has taken place, so it is not possible to point telescopes into the direction of the source beforehand, and observe the merger itself.

More detectors similar in size and targeted sensitivity to LIGO/Virgo are being constructed in India (LIGO India) and commissioned in Japan (KAGRA) \cite{KAGRA}, allowing for better estimates of the parameters of the detected black-hole binaries \cite{VeEA2012}. Building more detectors of the same kind can, however, improve the amount of information only so much. To improve our knowledge significantly, a new generation of detectors that is sensitive in other frequency bands is necessary. For these reasons, there is great interest in the potential of next-generation detectors like the Einstein Telescope (ET) \cite{PuEA2010} and the Laser Interferometer Space Antenna (LISA) \cite{ASEA2017}, and in the benefits of combining information from detectors in different frequency bands. Sesana investigated the prospects of multiband detections of stellar-mass BBHs in LISA and ground based-detectors \cite{Ses2016}. Moore et al. provided a detailed analysis of the detectability of stellar-mass BBHs in LISA \citep{LISAstellardetectability}. Isoyama et al. investigated what can be learned from multiband parameter estimation with B-DECIGO and ground-based detectors \cite{IsEA2018}, albeit without taking into account the changing detector orientation. Recent studies have investigated the capabilities of future detectors to constrain source parameters \cite{HaEv2019,AdEA2019}, and alternative theories of gravity \cite{TheoreticalPhysicsImplications, MiEA2010, Vit2016, ParametrizedPostNewtonian}.

Future detectors like ET and LISA, and possibly the first stage of the Deci-Hertz Interferometer Gravitational wave Observatory (B-DECIGO) \cite{NaEA2016}, will doubtlessly provide exquisitely precise information about astrophysical objects in our universe. Space-borne detectors sensitive at lower frequencies will provide year-long observations of individual signals, which will pose new challenges for data analysis. Changing detector orientation and position over the observation time is a complex problem. With certain approximations, it can be approached in frequency domain \cite{fdomaindetectormovement}. In this work, we simulate signals from compact binaries and the effects of detector motion in time domain, and subsequently perform a Fourier transformation to obtain the frequency-domain signal.

The objective of this work is to make predictions about the scientific gains we can expect from networks of next-generation detectors covering multiple disconnected bands or forming a broadband network, and to determine which detectors are best suited for a given science goal. To this end, we also investigate how the combination of data from different networks constrains source parameters. 

The structure of this work is as follows: Section \ref{sec:Detectors} presents the detectors we consider in our analysis, and how we model them in our simulation. In section \ref{sec:Formalism}, we will summarize the Fisher-matrix formalism for data analysis. The more technical aspects of our analysis, like the waveforms we use, the computation of the Fourier transform, etc. are covered in section \ref{sec:framework}. Section \ref{sec:lightBBHs} is devoted to the parameter estimation of stellar-mass, black-hole binaries with masses of the individual black holes in the range of $5-60 \msolar$. Intermediate-mass black holes (IMBHs) are black holes in the mass range $150 \msolar-500 \msolar$. Despite the fact that it is not clear if these black holes exist, and if they do, what their spatial and mass distribution is, we will explore how well we can reconstruct their properties with present and future detector networks in section \ref{sec:IMBH}. In section \ref{sec:NSbinaries}, we will see how the study of neutron-star binaries can benefit from future detectors.

\section{Detectors}
\label{sec:Detectors}

\begin{figure}[ht]
\centering
\includegraphics[width=0.95\columnwidth]{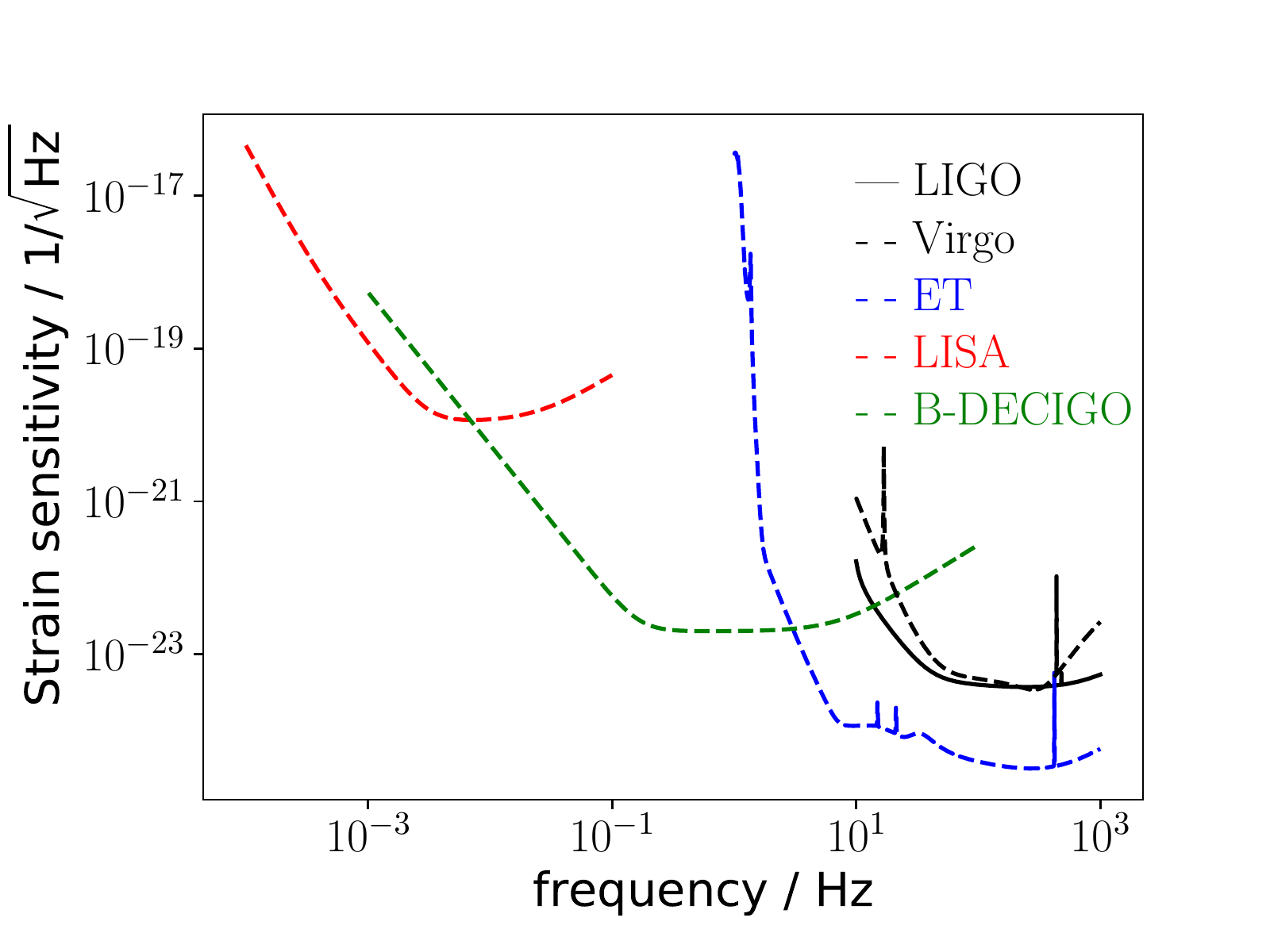}
\caption{Sensitivity curves of the detectors we use for our simulation.}
\label{fig:noisecurves}
\end{figure}

In this section we present the detectors we include in our simulation, and the assumptions we make when simulating them. Figure \ref{fig:noisecurves} shows the sensitivity curves we use for the different detectors.

\subsection{LIGO/Virgo}

The currently operating detectors of the LIGO and Virgo collaborations are sensitive in the frequency range from 10\,Hz up to about 1000\,Hz. For the LIGO detectors, we use the sensitivity curves from \cite{LIGOnoise}. For the Virgo detector, the sensitivity curve is the one from \cite{Virgonoise}. We simulate the detectors at their respective sites. As usual, the estimation of source positions is enhanced by combining data from detectors forming a network. More details about that in section~\ref{sec:Formalism}.

\subsection{Einstein Telescope (ET)}

The Einstein Telescope (ET) is a future ground-based gravitational wave detector. Its frequency range will be significantly wider than the one of current-generation detectors. It will extend from 3\,Hz to a few kHz. ET is characterized by a triangular shape with a total of three pairs of interferometers \cite{HiEA2009}. Each pair consists of a low-frequency and a high-frequency interferometer forming a so-called xylophone configuration, which will yield a sensitivity that is much better than the sensitivity of current-generation detectors and will allow us to see BBHs and neutron stars to much larger distances than before \cite{HaEv2019}. We represent each xylophone combining the low and high-frequency interferometers into a single sensitivity model (known as ET-D). The full ET triangle is then simulated as the sum of three xylophones with an arm-opening angle of $60\,\deg$, which are rotated against each other by $60\,\deg$. For the simulation of ET we use the noise curve from \cite{ETnoise}. The site of the future detector is not yet decided, but we use the coordinates of one of the proposed sites, which is Sardinia. The coordinates we use are N\ang{40;46.00;} E\ang{09;45.00;}.

\subsection{Laser Interferometer Space Antenna (LISA)}

The Laser Interferometer Space Antenna (LISA) is a future space-borne, gravitational-wave detector. It is designed to be sensitive at very low frequencies at about $10^{-4}-10^{-1}$\,Hz \citep{LISAmissionproposal}. Scientific goals include the detection of mergers of (super)massive black-hole binaries (SMBHs), galactic white-dwarf mergers, and extreme mass-ratio inspirals (EMRIs). The anticipated begin of the mission is in the 2030's. LISA is planned to consist of three spacecrafts orbiting the Sun on Earth-like orbits (trailing the Earth by approximately $20\,\deg$). The three spacecrafts form a triangle with an armlength of 2.5 million kilometers, and their orbits are slightly eccentric and tilted with respect to the orbit of the Earth, which leads to a rotation of the triangle in time (cartwheel motion) and temporal variations in arm length (breathing) \cite{TEA2004}. In our simulation, we use the position vectors of the spacecraft given in \cite{HaEA2008} to compute the unit vectors of the detector arms and the phase difference due to the detector position. Cutler showed in \cite{Cutler} that LISA can be simulated as the sum of two interferometers, which are rotated relative to each other by $45\,\deg$, and we follow his approach.

The sensitivity curve we use is the one from \cite{LISAsensitivity}. Since stellar-mass compact binaries emit for a very long time in the LISA band (for stellar-mass BBHs up to hundreds of years), and the LISA mission time is limited, the SNR and parameter estimate also depends on what part of the GW emission of the source is observed in LISA. Since we are mainly interested in multiband parameter estimation, and want to make sure that it does not take too long until the binary chirps to coalescence after the LISA detection, we choose our observation time such that it includes the last 3 years before the binary leaves the LISA frequency band at $0.1\,\text{Hz}$.

We work in the long-wavelength approximation and neglect effects arising from the frequency-dependent detector response. Generally speaking, the frequency-dependent detector response plays a role in LISA and has to be taken into account. For the cases investigated in this study, however, we have tested the ramifications of neglecting it, and it is a valid approximation. We include LISA only in the study of IMBH binaries, which accumulate the largest part of their SNR in the intermediate-to-low frequency range considering 3 years total observation time. So, for the purpose of parameter estimation, the high-frequency sensitivity model of LISA is not important in this study. For stellar-mass binary black holes merging within 3 years, which would spend more time emitting in the high-frequency end of LISA's observation band, the LISA high-frequency sensitivity model would be more important; but it does not change the fact that the SNR of these binaries is too low to assume that a significant number of them will be detected in LISA (see section~\ref{sec:lightBBHs}). So, we do not include LISA in the studies related to stellar-mass binary black holes.

\subsection{Deci-Hertz Interferometer Gravitational wave Observatory (B-DECIGO)}

The Deci-Hertz Interferometer Gravitational wave Observatory (DECIGO) is a proposed space-based gravitational-wave detector. Being sensitive from LISA's highest frequencies of around $10^{-1}$\,Hz up to 10\,Hz, it would close the sensitivity gap between LISA and the ground-based detectors. B-DECIGO is the first stage of the detector. It is planned to operate with a sensitivity that is somewhat lower than the sensitivity of the final detector. For the sensitivity curve, we use the fitting function given for the final DECIGO detector given in \cite{YaSe2011}, and scale it to the target sensitivity of B-DECIGO, which is $2 \times 10^{-23}\,\text{Hz}^{-1/2}$ \cite{NaEA2016}. Similarly to LISA, we simulate B-DECIGO as the sum of two interferometers rotated relative to each other by $45\,\deg$. At the cost of increased complexity, B-DECIGO could also be operated as three independent detectors, similar to ET, but so far this is not the plan.

The orbit of the B-DECIGO detector has not been decided yet \cite{Sato}. Among the proposed orbits is an orbit around the Sun similar to LISA, as well as several Earth-centered orbits (geostationary and others). For our simulation, we use an orbit around the center of the Earth with a radius of $36000\,\text{km}$ and a period of one day. The orbit is in the plane of the orbit of the Earth around the Sun. Regarding orbits around the Earth, the plane of the orbit, as well as its exact period and radius, should not have a large influence on parameter estimation.
We will see later, however, that it does make a difference if the orbit is LISA-like, as compared to Earth-centered. This is due to the vastly different time scale of the detector movement (the time scale for LISA-like orbits is a year, as compared to approximately a day for Earth-centered orbits). For this reason, we will also perform a parameter estimation for B-DECIGO with a LISA-like orbit, to see how it compares with an Earth-centered orbit.

Due to the fact that the armlength of B-DECIGO is $100\,\text{km}$, much shorter than the armlength of LISA, the detector response is not frequency-dependent, and the long-wavelength approximation is justified.

Stellar-mass BBHs and intermediate-mass BBHs spend less than 3 years in the B-DECIGO band. Only some neutron stars spend more than three years in B-DECIGO. We limit the maximum observation time in B-DECIGO to the last 3 years before the binary leaves the B-DECIGO frequency range at 10 Hz.

\section{Fisher matrix formalism}
\label{sec:Formalism}

The Fisher-matrix approach is a widely used, and computationally inexpensive method for parameter estimation. The method is based on an expansion of the likelihood around the true source parameters. The limitations of its applicability have been investigated in \cite{Val2008,RoEA2013}. In this section, we follow \citep{Val2008,FisherSkyLoc}. The detector data is understood as a sum of instrument noise $n$ and a signal $h(\vec\theta_0)$ that is described by a model with true parameters $\vec\theta_0$:
\begin{align}
s = h(\vec\theta_0) + n,
\end{align}
Model parameters estimated from an observation $s$ generally deviate from the true parameters. The noise-weighted inner product is defined as
\begin{align}
\label{eq:scalarproduct}
(a,b)=4 \, \Re \, \int\limits_0^\infty \frac{a(f) b^*(f)}{S_n(f)} \dd f.
\end{align}
The likelihood of observing the data $s$ when we have a binary with parameters $\vec \theta_0$ can be expressed as
\begin{align}
p(s|\vec\theta_0) \propto \exp\left[-\left(s-h(\vec\theta_0),s-h(\vec\theta_0)\right)/2\right].
\end{align}
Now we expand the signal model around the true source parameters $\vec\theta_0$:
\begin{align}
h(\vec\theta\,) = h_0 + \Delta \theta_k \partial_k h+ \Delta \theta_k \Delta \theta_j \partial_k \partial_j h + ...
\end{align}
Here $\Delta \theta_k$ denotes the deviation of model parameter $\theta_k$ from the true values $\theta_{0,k}$, and $\partial_k h = \partial_k h(\vec\theta\,)|_{\vec\theta=\vec\theta_0}$ is the derivative of the signal at the true values with respect to parameter $\theta_k$. $h_0\equiv h(\vec\theta_0)$ is the true signal.
The matrix with the entries
\begin{align}
\label{eq:FIM}
F_{kj} = (\partial_k h, \partial_j h)
\end{align}
is the \textit{Fisher-matrix.} By making use of the above expansion of the likelihood, it can be shown \cite{Val2008} that in the high-SNR limit, the covariance matrix can be approximated as the inverse of the Fisher matrix:
\begin{align}
\langle\theta_k \theta_j\rangle \approx (\partial_k h, \partial_j h)^{-1}.
\end{align}
This allows us to compute standard deviations,
\begin{align}
\label{eq:error}
\langle (\Delta\theta_i)^2 \rangle^{1/2} \approx \sqrt{(\partial_i h, \partial_i h)^{-1}}.
\end{align}
The SNR can be computed as
\begin{align}
\text{SNR} = \sqrt{4\int\limits_0^\infty \frac{|h(f)|^2}{S_n(f)} \dd f} = \sqrt{(h, h)}.
\end{align}

\section{Analysis framework}
\label{sec:framework}

In this section, we discuss the more technical problems that have to be addressed to employ the Fisher-matrix approach. We use the waveform model for the two gravitational-wave polarizations $H_+$ and $H_\times$ of the gravitational-wave tensor from \cite{HaEA2008}:
\begin{align}
\label{eq:pluspertrubation}
H_+(t) = \frac{c}{2 r} \left[5 \frac{\Mc^5}{\tc-t}\right]^{1/4} (1+\cos^2(\iota)) \cos(\phi(t) + \phi_\text{c})
\end{align}
and
\begin{align}
\label{eq:crosspertrubation}
H_\times(t) = \frac{c}{2 r} \left[5 \frac{\Mc^5}{\tc-t}\right]^{1/4} 2 \cos(\iota) \sin(\phi(t) + \phi_\text{c}).
\end{align}
The time dependence of the phase $\phi(t)$ is
\begin{align}
\phi(t) = - \frac{2}{\eta} \sum_{k=0}^7 p_k \tau^{(5-k)/8},
\end{align}
with the coefficients $p_k$ as given in \cite{HaEA2008}.
We have $\tau=(\eta c^3 (\tc-t))(5 G \Mtot)$, and we used the symmetric mass ratio $\eta = \mu/\Mtot$.

When looking at the time dependence of the phase, we can see an important feature of the waveform that will become important later. For times far away from merger, $\tc-t \gg G\Mtot/c^3$, we can approximate the time dependence of the phase with the term with the largest power-law exponent:
\begin{align}
\phi(t) = - \frac{2}{\eta}  \tau^{5/8} \propto \Mc^{-5/8} (\tc-t)^{5/8}.
\end{align}
Here we see that far away from merger, the phase only depends on the chirp mass $\Mc=\eta^{3/5} \Mtot^{2/5}$, which is a function of the two mass parameters. That means that we can obtain the same phase evolution with an infinite number of combinations of the two mass parameters $\eta$ and $\Mtot$; the two parameters are degenerate. The amplitude only depends on the chirp mass as well. For this reason, we expect that space-borne detectors, which detect the waveform far away from merger, will provide precise information about the chirp mass. The estimate of the individual mass parameters will, however, be limited by their degeneracy in the early inspiral.

It is also important to note that this waveform is valid in the inspiral only. For this reason, we suppress it with a tempering function when it reaches the frequency of the innermost stable circular orbit, and we do not have a model for the merger part of the waveform. This means that the information contained in the merger part of the waveform is lost, and our estimates for the standard deviations will be larger than the ones we could theoretically obtain if we had a complete waveform. Similarly, our analysis will systematically underestimate the SNR in detectors that would be able to detect the merger waveform - typically LIGO/Virgo and ET, but for IMBHs also B-DECIGO (some IMBHs merge in the B-DECIGO frequency range). Still, the truncated waveform is sufficient for our purposes. What we want to provide is a qualitative prediction of the relative performances of the different detectors, and of the benefits of combining them; to this end, the waveform we use is sufficient.

Now we can construct the metric perturbation as
\begin{align}
H =
\begin{pmatrix} 
H_+ & H_\times & 0 \\
H_\times & -H_+ & 0 \\
0 & 0 & 0
\end{pmatrix}
\end{align}
In the next step, we multiply the metric perturbation with a rotation matrix $R(\theta,\phi,\psi)$. This matrix represents a sequence of Euler rotations \cite{Val2005a}, which is a function of the position vectors $\theta$ and $\phi$ of the source and the polarization vector $\psi$ of the wave. We obtain
\begin{align}
H_\text{det} = R \, H \, R^T.
\end{align}
This is necessary to rotate the gravitational wave into the detector frame; $H_\text{det}$ denotes the metric perturbation in the detector frame. Finally, the metric perturbation is projected onto the unit vectors $\vec{e_1}$, $\vec{e_2}$ of the detector that is under consideration, to obtain the strain $h$ in the detector:
\begin{align}
h = \vec{e_1}^T H_\text{det} \vec{e_1} - \vec{e_2}^T H_\text{det} \vec{e_2}
\end{align}
With the waveform we use, it is easy to calculate derivatives analytically. The derivatives of the phase $\phi(t)$ with respect to the mass parameters and the coalescence time are given in \cite{HaEA2008}.
For the parameters which appear in the waveform itself, we simply take the derivative of the metric perturbation $H$
\begin{align}
\partial_i H =
\begin{pmatrix} 
\partial_ i H_+ & \partial_ i H_\times & 0 \\
\partial_ i H_\times & - \partial_ i H_+ & 0 \\
0 & 0 & 0
\end{pmatrix},
\end{align}
multiply with the rotation matrices and project onto the detector arm unit vectors to obtain the derivative of the detector strain $\partial_i h$.
For the position angles $\theta$, $\phi$ and the polarization angle $\psi$, which do not appear in the waveform itself, but in the rotation matrices, we have to take the derivative of the rotation matrices and apply the chain rule
\begin{align}
\partial_i H_\text{det} = (\partial_i R) \, H \, R^T + R \, H \, (\partial_i R^T),
\end{align}
before we can project onto the detector as before. When considering  a network of detectors, the Fisher matrix element is the sum of the contributions of the individual detectors,
\begin{align}
F_{kj} = \sum_{\text{detectors}} (\partial_k h, \partial_j h).
\end{align}
The SNR is
\begin{align}
\text{SNR} = \sqrt{\sum_{\text{detectors}}(h,h)}.
\end{align}

At this point, it is necessary to discuss how source-triangulation information enters the Fisher matrix. As an example, when we consider the network of the LIGO/Virgo detectors, the gravitational wave will be detected earlier in the detector that is closest to the source. We can express this by saying that there is a source position-dependent time shift
\begin{align}
\Delta t = \vec{n} (\theta,\phi) \, \vec{r}_\text{det}/c
\end{align}
($\vec{n}(\theta,\phi)$ is the unit vector pointing to the source, $\vec{r}_\text{det}$ is the position vector of the detector). We follow \cite{FisherSkyLoc} and define the arrival time as the time when the gravitational wave arrives at the center of the Earth. Then, the arrival times in the detectors are $t_\text{c} - \Delta t$. Since the time shift depends on the position angles $\theta$ and $\phi$, we have to include an additional term in the derivatives of the detector strain with respect to the source position parameters. The complete derivative of the metric perturbation with respect to the extrinsic source parameters is then
\begin{align}
\partial_\alpha H_\text{det} = (\partial_\alpha R) \, H \, R^T + R \, H \, (\partial_\alpha R^T) + R \, \left(\partial_\alpha H \right) \, R^T.
\end{align}
Here $\alpha$ represents the extrinsic source parameters $\theta$ and $\phi$.

After performing these steps, we have the derivative of the detector signal $\partial_i h$ in time domain. A fast Fourier transform (FFT) gives us the derivative of the detector signal in frequency domain $\partial_i \tilde{h}(f)$, which allows us to compute the Fisher matrix. After inverting the Fisher matrix, we can compute the standard deviation of the parameter estimate with eqs. \eqref{eq:scalarproduct}, \eqref{eq:FIM},  and \eqref{eq:error}.

\begin{figure}[t]
\centering
\includegraphics[scale=0.5]{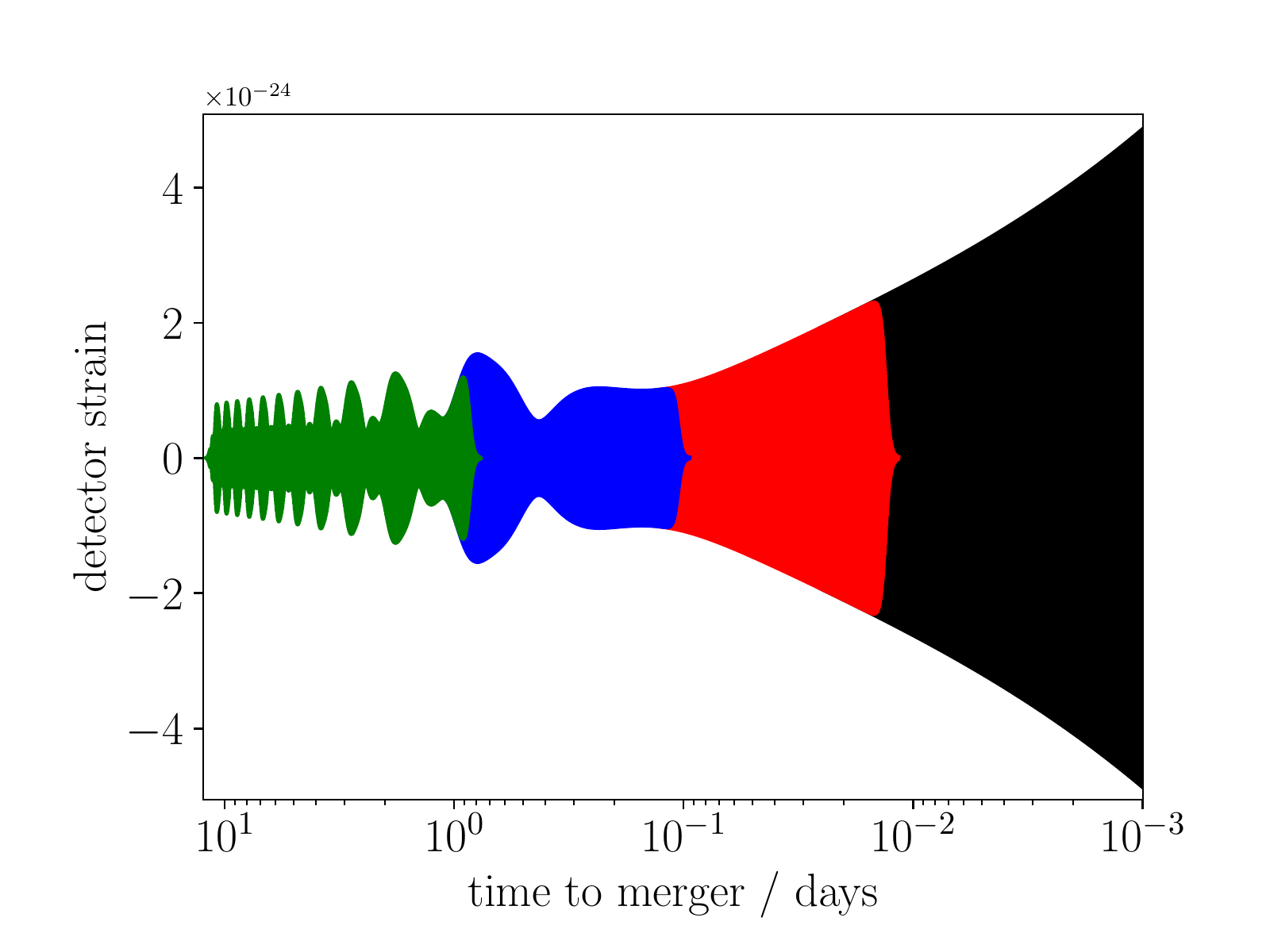}
\caption{Time series of the signal of a neutron star merger in ET. Colors show the different time segments.}
\label{fig:NStimeseries}
\end{figure}

Some more comments are due with respect to the unit vectors of the detectors and the FFT. Concerning the detector projection, we use time-dependent detector arm vectors and detector positions. This takes into account the rotation of the Earth for the ground-based detectors (LIGO/Virgo and ET) and the movement of the spacecraft for LISA and B-DECIGO. In general, the detector movement affects the signal in the detector in two ways. First, the strain in the detector depends on the orientation of the detector arms, which is time dependent. This leaves an imprint on the amplitude of the detector strain, which reflects the periodicity of the movement (daily rotation for detectors on Earth, yearly rotation for space-based detectors orbiting the Sun). This can be seen in fig.~\ref{fig:NStimeseries}, which shows the signal of a neutron star binary merger in ET. Note that the time axis is log-scaled, and the source emits gravitational waves with frequencies above 1 Hz for several days. It is interesting to see how the daily modulation of the signal amplitude appears in the time domain signal at time scales comparable to a day and longer. As a result, the FFT, which is shown in fig.~\ref{fig:NSFFTfinal}, is suppressed at frequencies at which the detector orientation is such that the signal is weak. Second, due to the orbit of the detector around the Sun (for both ground- and space-based detectors), the detector-source distance changes, which leads to an additional phase in the waveform (Doppler effect). For the LIGO/Virgo detectors, the signal duration is so short that the detector can be approximated as stationary, and these effects can be neglected. In the space-borne detectors LISA and B-DECIGO, however, signals will be observed for months or years, and movement effects have to be taken into account and have important ramifications for parameter estimation. ET stands between these two cases;  a $10+10\,\msolar$ BBH emits frequencies above 2 Hz for 41 minutes, heavier BBHs spend less time in this frequency range. In these cases movement effects can also be neglected. Neutron-star binaries, on the other hand, can spend more than a day in the ET frequency band, so here movement effects are important. This means that in ET, it depends on the astrophysical object that is being observed: the lighter the object, the longer the observation time.

\begin{figure}[t]
\centering
\includegraphics[scale=0.5]{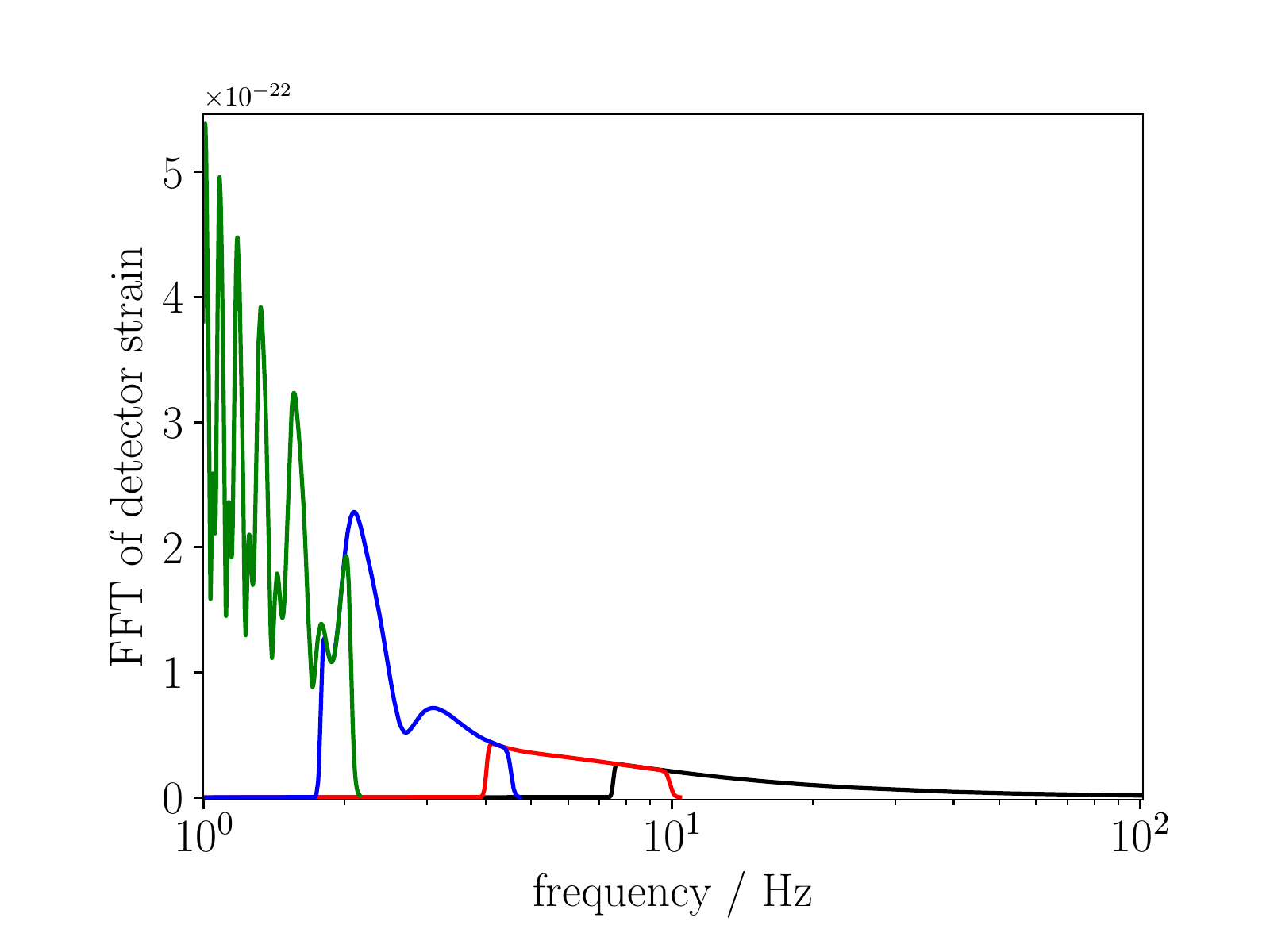}
\caption{Fast-Fourier transform (FFT) of the time-domain signal of a neutron star merger in ET. Colors show the different time segements (color code is identical to fig.~\ref{fig:NStimeseries}).}
\label{fig:NSFFTsegments}
\end{figure}

\begin{figure}[t]
\centering
\includegraphics[scale=0.5]{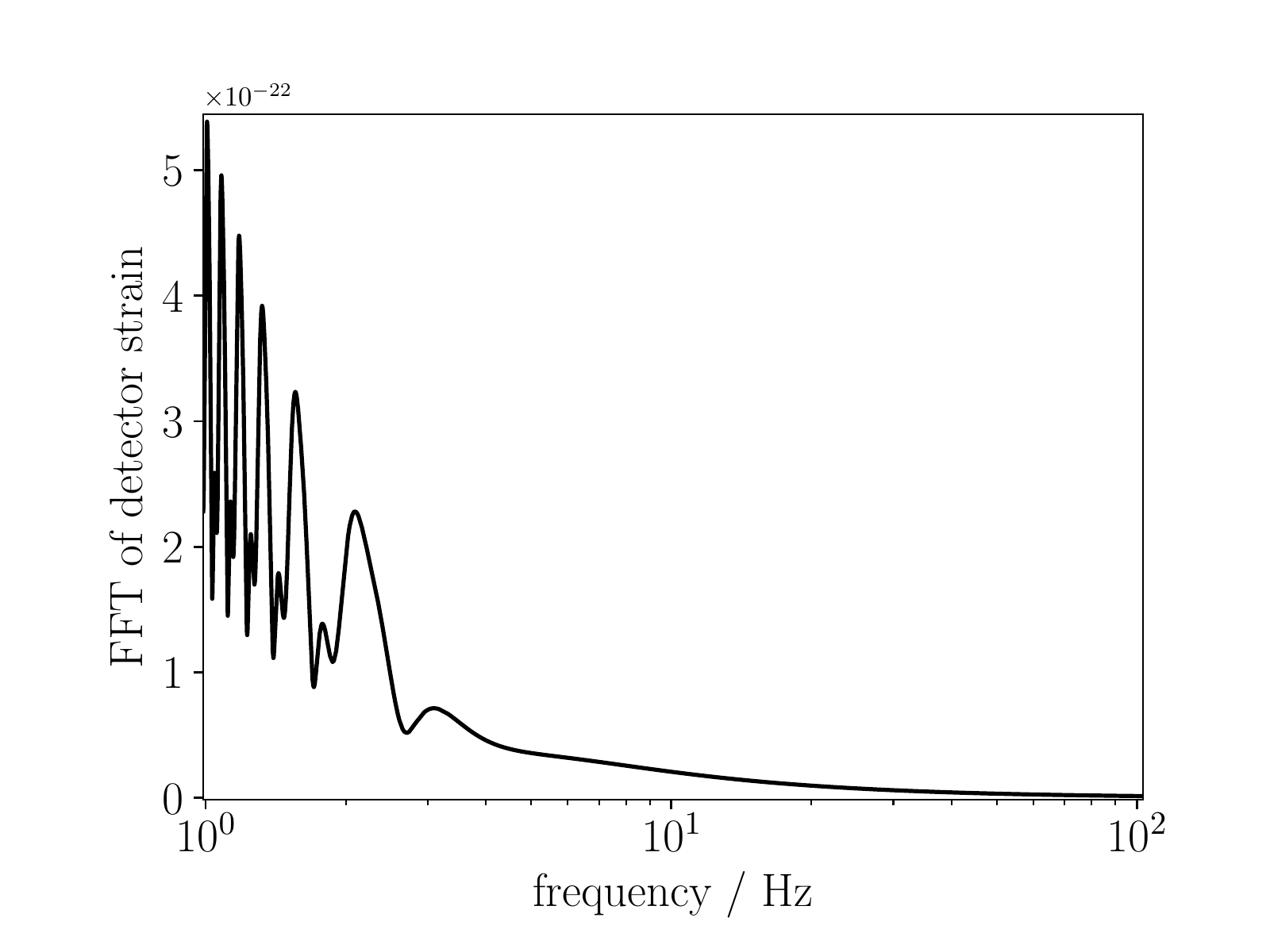}
\caption{Final fast-Fourier transform (FFT) of the time-domain signal of a neutron star merger in ET.}
\label{fig:NSFFTfinal}
\end{figure}

Also with respect to the computation of the Fourier transform, the frequency band of the detector and the astrophysical object that is being studied play an important role. It is computationally expensive to simulate signals over long periods of time and with a high time resolution. This does not pose a problem in the LIGO/Virgo sensitivity band, in which we have to compute signals of a length of a few seconds and typical frequencies do not exceed a few $100\,\text{Hz}$. The LISA band is computationally more challenging, since we are dealing with observation times of years, but as the maximum frequency is only $0.1\,\text{Hz}$, it can still be done. It is in ET and B-DECIGO where the situation is more challenging. These detectors combine long observation times with relatively high frequencies, i.e., there observation bands are broader (on logarithmic scale) than of other detectors. When observing a neutron star binary merger in ET, we have an observation time of more than a day and a maximum frequency in the kHz range. In B-DECIGO, the maximum frequency is lower, around 10 Hz, but the observation time is longer, several months, which makes the simulation in the time domain and the subsequent FFT equally problematic. The point is that in both cases the high time resolution is necessary only in a very small part at the end of the time domain signal (close to the merger), and for the rest of the observation time a lower time resolution would suffice, since the frequency drops very quickly as we move away from the merger. The solution to this problem is that we split the observation time into several segments, which are shortest close to the merger and get progressively longer further away from the merger. Similar approaches to reduce computational costs have been developed for the waveform generation of long signals for parameter estimation in advanced 2G and 3G detectors \cite{MBTA1,MBTA2}. Figure \ref{fig:NStimeseries} shows the signal of a neutron-star binary merger in ET, and the segments of the time series are shown in different colors. The algorithm we use is the following: We start from the maximum frequency of the gravitational waves observed in the detector, $f_\text{max}$, and the minimum frequency of the detector, $f_\text{min}$. Then we create $n_\text{frames}-1$ frequencies between these two frequencies, spaced logarithmically ($n_\text{frames}$ is the number of time frames we use). For example, if we have $n_\text{frames}=4$ time frames, we have the frequencies $f_\text{max} = f_0>f_1>f_2>f_3>f_4 = f_\text{min}$. Since we know the phase evolution $\phi(t)$ of the gravitational waveform we use, we can easily compute the frequency $f = 1/(2\pi)\dd \phi(t)/\dd t$, at which a binary emits at time $t$, and we can also solve the inverse problem, namely computing the time at which a binary emits gravitational waves of a certain frequency. This allows us to easily compute the length of the time frames we need, $T_i = t(f_{i})-t(f_{i+1})$. We make the time frames a bit longer than that, to have an overlap between neighboring frames. In each time frame $T_i$, the highest frequency that occurs is the frequency $f_i$, and we choose the time resolution we need to resolve this frequency, which corresponds to a sampling frequency $2 f_i$. We simulate the time-domain signal (or its derivative, for the Fisher matrix) in each of the time frames, and perform the FFT.  At the beginning and the end of each time frame, we have to suppress the time-domain signal with a tempering function, to avoid artifacts - we can do that because the time segments overlap a bit. Then we combine the FFTs of the different segments into a final FFT, which is the one we use. Fig.~\ref{fig:NSFFTsegments} shows the FFTs of the different segments (color code is identical to fig.~\ref{fig:NStimeseries}), and fig.~\ref{fig:NSFFTfinal} shows the final FFT. This is possible because the frequency of the gravitational waves detected at a certain time $t$ is a linearly growing function of $t$. Therefore, there is a one-on-one correspondence of time to frequency $t \Leftrightarrow f$. This means that each time segment contributes to a certain frequency interval of the FFT, and there are no correlations between the FFTs, so we can easily combine them.

\section{Stellar-mass black hole binaries: LIGO/Virgo, ET, and B-DECIGO}
\label{sec:lightBBHs}

In this section, we investigate BBHs which consist of stellar-mass black holes in the range of $5-60 \msolar$. In its currently planned configuration, LISA can only detect these binaries with sufficient SNR if they are very close (luminosity distance $<200 \Mpc$) and rather heavy (masses of both black holes $>40 \msolar$); see also fig.~3 in \cite{ASEA2017}. If we require that the binaries chirp to coalescence in less than 3 years, such that we can perform a multiband parameter estimation, there will be very few candidate binaries. None of the BBHs from the population used in this paper was detectable by LISA (even increasing to a population of size 10$^5$ BBHs does not lead to a single detection). These findings are consistent with the results presented in \cite{LISAstellardetectability}, which show that stellar-mass BBHs are on the border of detectability in LISA; even under favorable circumstances, only few events are expected. Under less favorable circumstances, there may be no detections at all.

B-DECIGO, on the other hand, can detect stellar-mass BBHs up to large distances. Since B-DECIGO is sensitive at higher frequencies than LISA, binaries observed in B-DECIGO also chirp into the frequency range of ground-based detectors in a sufficiently short time. For this reason, the following analysis is limited to the LIGO/Virgo detectors, ET, and B-DECIGO.

\subsection{Detection capabilities of the detectors}

\begin{figure}[t]
\centering
\includegraphics[scale=0.5]{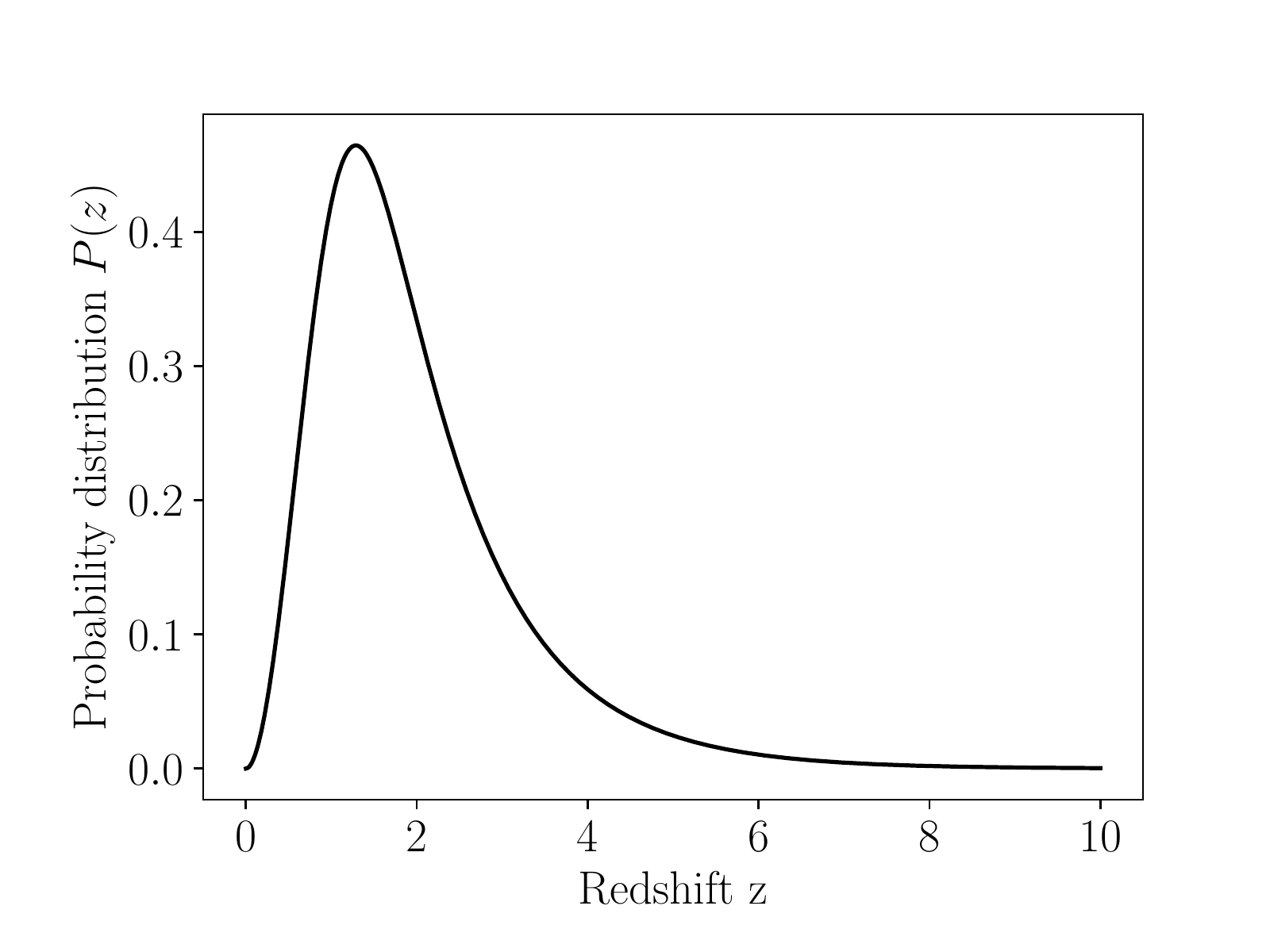}
\caption{Redshift-dependence of the spatial probability distribution we use for our cosmological BBH distribution.}
\label{fig:Pz}
\end{figure}

\begin{figure}[t]
\centering
\includegraphics[scale=0.5]{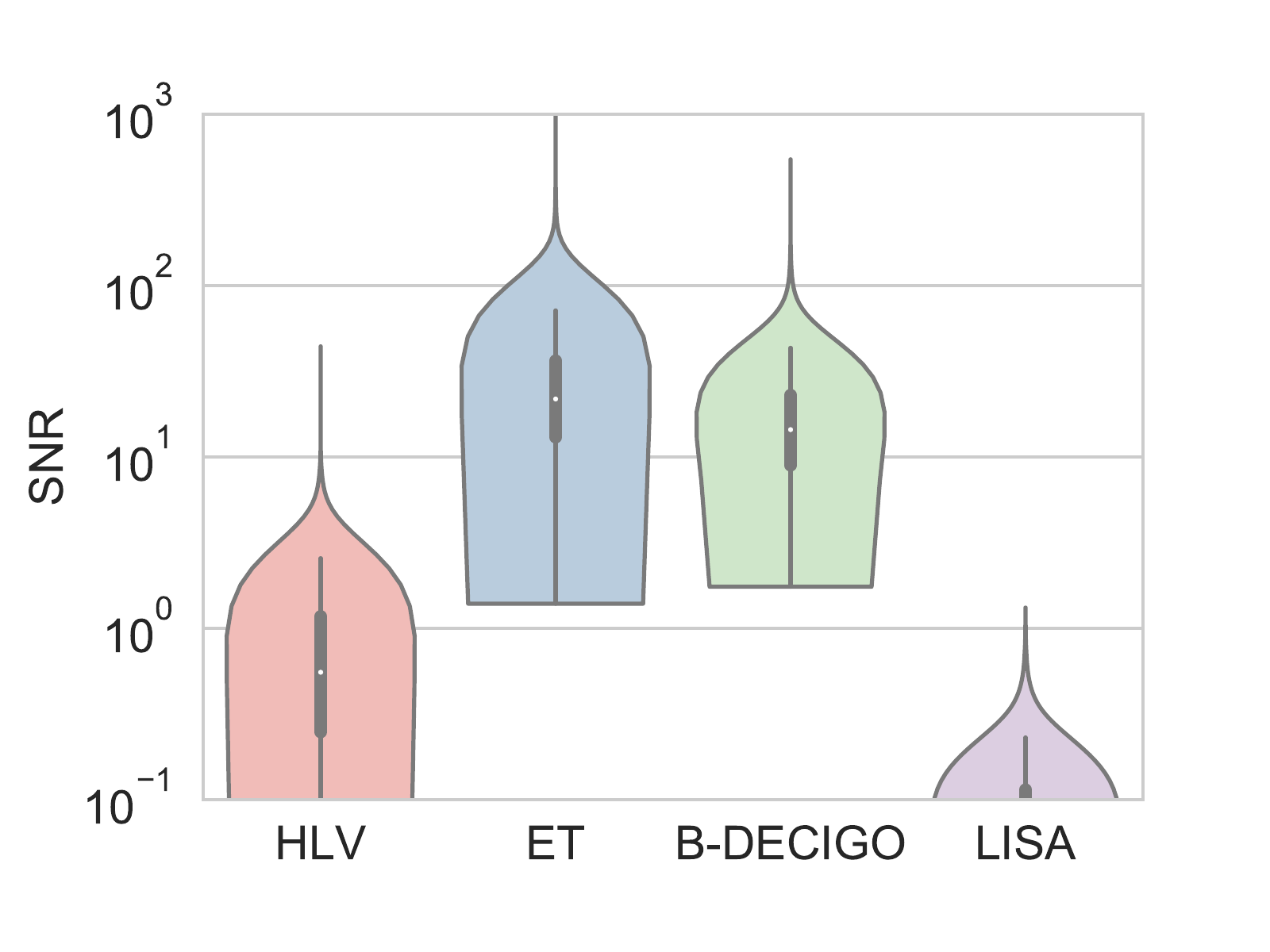}
\caption{Distribution of SNRs in the HLV detectors, ET and B-DECIGO, for a cosmological distribution of 1000 stellar-mass binary black holes.}
\label{fig:SNRdist1}
\end{figure}

\begin{table}
\begin{tabular}{c|c|c|c|c}
$\text{SNR}>10$ events & HLV & ET & BDEC & LISA \\ 
\hline 
out of 1000 events & 1 & 846 & 697 & 0 \\ 
\end{tabular}
\caption{Number of events with a $\text{SNR}>10$ in the different detectors, out of a cosmological distribution of 1000 stellar-mass binary black holes.}
\label{table:numberofevents1}
\end{table}

As a first step, we want to understand what the detection capabilities of the detectors under consideration are. The black holes detected so far by the LIGO and Virgo detectors are consistent with a mass probability distribution with a power-law index of $1.6$ \cite{LVC2019}, and recent studies indicate that the mass distribution depends only mildy upon the redshift \cite{MaEA2019}. Therefore, we assume that the mass probability distribution of the black holes is a power law with an index of $1.6$, and that the mass distribution is independent of redshift. For the redshift dependence of the spatial probability distribution, we use the model used in \cite{redshiftdependence}. It is shown in fig.~\ref{fig:Pz}.

With these assumptions, we generate a cosmological distribution of 1000 black-hole binaries; the masses of the black holes are in the mass range $5 \msolar<m_\text{BH}<60 \msolar$. Fig.~\ref{fig:SNRdist1} shows the distribution of SNRs of these binaries in the detectors. We can see that the future detectors ET and B-DECIGO will be much more sensitive than the currently operating detectors, with ET achieving higher SNRs for stellar-mass BBHs than B-DECIGO. For a detection we require $\text{SNR}>10$. Table~\ref{table:numberofevents1} shows the numbers of the events with $\text{SNR}>10$. We can see that both ET and B-DECIGO can detect the vast majority of the BBHs (846 and 697 out of 1000). The LIGO/Virgo detectors, according to their design sensitivity, which they have not reached yet, are much less sensitive and detect only 1 out of 1000 BBHs. Any event that can be detected with the LIGO/Virgo detectors can also be detected with ET and B-DECIGO. Therefore we can conclude from this simulation that B-DECIGO is a detector that would allow for multiband parameter estimation. With a combination of B-DECIGO and ET, multiband parameter estimation with a much larger number of events would become possible. LISA as a low-frequency detector is not able to detect any of the stellar-mass BBHs considered here. For this reason, we do not include LISA in this parameter-estimation study of stellar-mass black-hole binaries.

\subsection{An example event}

In this section, we take a look at an example event to illustrate how multiband parameter estimation can work. The binary we are looking at consists of two black holes with masses $m_1 = 16 \msolar$ and $m_2 = 10 \msolar$, at a distance of $r = 400 \Mpc$, with inclination angle $\iota = 45\,\deg$. The SNR in the detectors is shown in table \ref{table:SNR1}. The estimates of standard deviations of the parameters in the different detectors are given in table~\ref{table:errors}.

\begin{table}
\begin{tabular}{c|c|c|c|c}
• & HLV & ET & BDEC & LISA \\ 
\hline 
SNR & 22 & 804 & 179 & 0.69 \\ 
\end{tabular}
\caption{SNRs of the stellar-mass BBH example binary in different detectors.}
\label{table:SNR1}
\end{table}

\begin{table*}
\begin{tabular}{c|c|c|c|c|c|c|c|c}
• & HLV & ET & BDEC  & BDEC2 & HLV/BDEC & ET/BDEC & HLV/ET  & HLV/ET/BDEC \\ 
\hline 
$\Delta \mu$ / $\msolar$ & $7.6 \times 10^{-2}$  & $5.8 \times 10^{-4}$ & $6.3 \times 10^{-4}$ & $8.2 \times 10^{-4}$ & $4.3 \times 10^{-4}$  & $7.3\times 10^{-5}$ & $5.4\times 10^{-4}$ & $7.3 \times 10^{-5}$ \\ 
\hline 
$\Delta \Mtot$ / $\msolar$ & $4.4 \times 10^{-1}$  & $3.4 \times 10^{-3}$ & $4 \times 10^{-3}$ & $5.2 \times 10^{-3}$ & $2.8 \times 10^{-3}$  & $4.6 \times 10^{-4}$ & $3.2\times 10^{-3}$ & $4.6 \times 10^{-4}$ \\ 
\hline 
$\Delta r$ / $\Mpc$ & $152$ & $23$ & $7.9$ & $13.3$ & $7.8$ & $2.4$ & $3.7$ & $2.4$ \\ 
\hline 
$\Delta \iota / \deg$ & $20$ & $0.6$ & $1$ & $1.6$ & $1$ & $0.29$ & $0.3$ & $0.29$ \\ 
\hline 
$\Delta \theta / \deg$ & $4.2$ & $43.8$ & $0.13$ & $0.15$ & $0.025$ & $0.022$ & $2$ & $0.022$ \\ 
\hline 
$\Delta \phi / \deg$ & $3.7$ & $11.6$ & $0.015$ & $0.033$ & $0.014$ & $0.013$ & $1.7$ & $0.013$ \\ 
\hline 
$\Delta \psi / \deg$ & $83$ & $7$ & $2$ & $3.3$ & $2$ & $0.58$ & $1.16$ & $0.58$ \\
\hline 
$\Omega / \deg^2$ & $38$ & $1390$ & $5.5 \times 10^{-3}$ & $7.3 \times 10^{-3}$ & $3.7 \times 10^{-5}$ & $1.8 \times 10^{-5}$ & $10.7$ & $1.8 \times 10^{-5}$ \\
\end{tabular}
\caption{Fisher estimates for the standard deviations of the stellar-mass example binary for different detectors. BDEC is B-DECIGO with a geostationary orbit, BDEC2 with a LISA-like orbit.}
\label{table:errors}
\end{table*}

The fact that the SNR in LISA is sub-detection threshold, even for this example binary, which is very close, clearly indicates that we cannot expect to perform multiband analyses of stellar-mass black hole binaries in LISA. We remind that we assume a maximal observation time of 3 years in the LISA band.

\begin{figure}[t]
\centering
\includegraphics[scale=0.5]{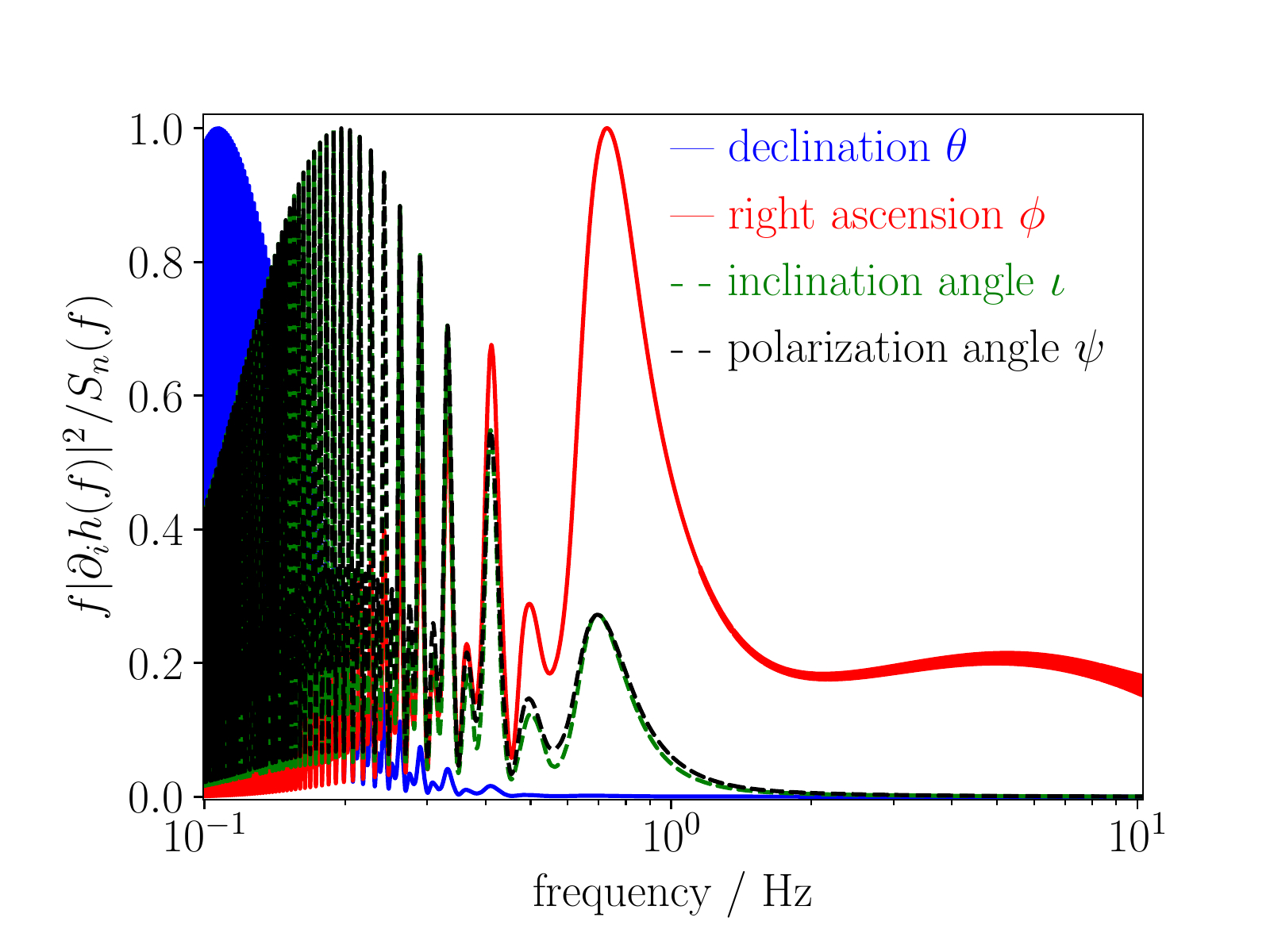}
\caption{Frequency-dependence of the information content of the diagonal element of the Fisher-matrix in BDEC (Earth-centered orbit).}
\label{fig:stellarBDECinfoplot}
\end{figure}
\begin{figure}[t]
\centering
\includegraphics[scale=0.5]{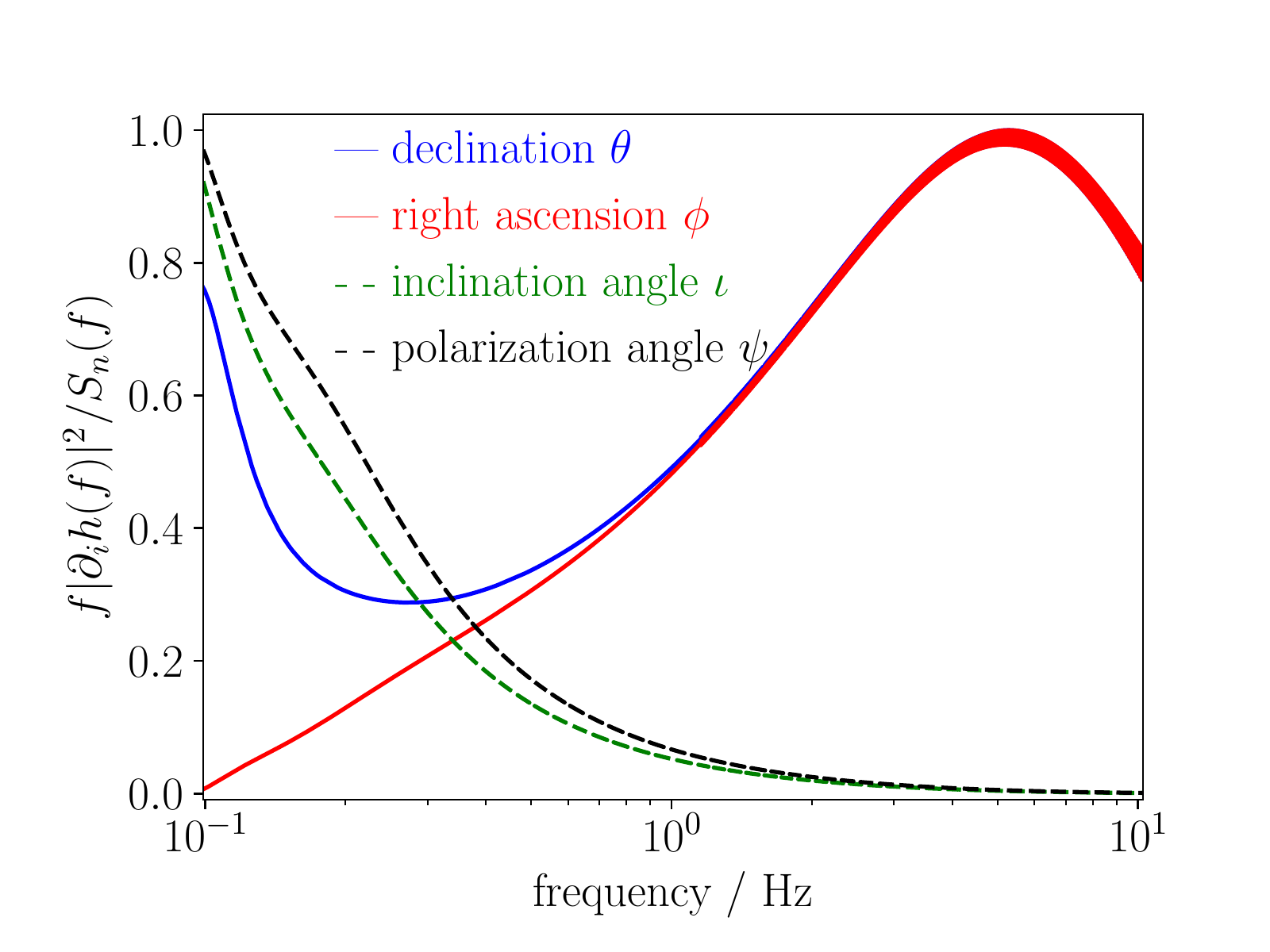}
\caption{Frequency-dependence of the information content of the diagonal element of the Fisher-matrix in BDEC2 (LISA-like orbit).}
\label{fig:stellarBDEC2infoplot}
\end{figure}

Table \ref{table:errors} shows the standard deviations of the parameter estimates of the LIGO/Virgo detectors (HLV), ET, and two possible realizations of the B-DECIGO detector (BDEC is the geostationary orbit, BDEC2 the LISA-like orbit), as well as combinations of these detectors. When considering the estimates of the mass parameters, we mean the detector-frame mass parameters. Comparing the two possible B-DECIGO realizations, we can see that in general, BDEC gives more precise estimates than BDEC2. What we can see here is the effect of differing time scales of the detector movement. The detector movement breaks the degeneracy between the parameters $\theta$, $\phi$, $\iota$, $\psi$, and $r$. This improves the estimate of these parameters, and it explains why BDEC generally gives better estimates. To understand this better, we plot in fig.~\ref{fig:stellarBDECinfoplot} and fig.~\ref{fig:stellarBDEC2infoplot} the function $f|\partial_i h(f)|^2/S_n(f)$, which is related to the contribution of a certain frequency to the diagonal element $(i,i)$ of the Fisher-matrix. We normalize each graph to the maximum value. It is important to note that this quantity is only indirectly related to the accuracy of the estimate of parameter $i$, since the off-diagonal elements also play a crucial role. Nevertheless, we can see in these plots that due to the detector movement, for different parameters, information about the parameter is accumulated at different frequencies. And we see that this effect is stronger for BDEC than for BDEC2 when comparing fig.~\ref{fig:stellarBDECinfoplot} and fig.~\ref{fig:stellarBDEC2infoplot}. In BDEC2, the two source position angles are completely degenerate at high frequencies. The degeneracy is only broken at lower frequencies, where detector movement effects begin to become important.

It is also clear that this effect depends on the mass of the binary we are looking at: heavier binaries will spend less time in the B-DECIGO frequency range (but still much more than a day), and the discrepancy between BDEC and BDEC2 will be larger, see section \ref{sec:IMBH}. Lighter astrophysical sources like neutron-star binaries, on the other hand, will spend more time in the B-DECIGO frequency range, and the discrepancy will be smaller, see section \ref{sec:NSbinaries}. Comparing BDEC to the other detectors, we see that it gives the most precise estimates, except for the mass parameters and the inclination angle, where ET gives more precise estimates. The mass parameters are measured very precisely in BDEC, and the sky localization is better than the sky localization of the HLV detector network. B-DECIGO also gives a precise measurement of the polarization angle $\psi$, the inclination angle $\iota$ and of the distance $r$. This is due to the fact that the long observation time and the movement of the detector break the degeneracy between $\theta$, $\phi$, $\psi$, and $\iota$, whereas the time delay triangulation of the LIGO/Virgo network only breaks the degeneracy of $\theta$ and $\phi$.
In ET, all parameters except the mass parameters and the inclination angle are measured much less precisely than in B-DECIGO. This is due to the fact that ET neither benefits from time triangulation (because it is only one detector) nor from detector movement effects (since the observation time is short). Due to the shorter observation time and the lower sensitivity, the estimate of the mass parameters is two orders of magnitude worse in the HLV detector network. The sky localization is worse than the one obtained from B-DECIGO. In comparison to B-DECIGO, the HLV detector network is not able to determine the polarization angle $\psi$, the inclination angle $\iota$ and the distance $r$ precisely, which is due to the parameter degeneracy and the absence of movement effects, as explained above.

Now let us take a look at the predictions we obtain when we combine the information from several detectors. With respect to the mass parameters, we can see that the parameter estimates of B-DECIGO improve significantly (by up to an order of magnitude) if B-DECIGO is combined with a ground-based detector (HLV or ET, or both). This is due to the feature of the waveform mentioned in section~\ref{sec:Formalism}: far away from merger, the mass parameters are highly degenerate (i.e. a certain change in the waveform can be achieved either by changing $\mu$ or by changing $\Mtot$), the estimate of the individual mass parameters is limited by this degeneracy. A ground-based detector that detects the waveform close to merger provides the necessary additional information to break the degeneracy and thereby improve the estimate. This effect is even stronger for space-borne detectors at lower frequencies, which is what we will see in LISA when considering IMBHs. The combination of ET and HLV gives an estimate that is very close to the estimate of ET alone, so in this case, there is no significant improvement. With respect to the sky localization, we can see that combining ET with the HLV network gives the largest improvement of the error estimate, since it allows for time triangulation with four detectors. The determination of $\iota$ and $\psi$ also benefits from combining HLV with ET.

The most important advantage of B-DECIGO, which is not contained in table \ref{table:errors}, is that B-DECIGO can observe a binary over months, and it can determine the position of the source before the merger takes place. This will make it possible to point telescopes at the source when the merger takes place, and to capture the electromagnetic counterpart. Combining the gravitational wave detection with the electromagnetic counterpart will allow for detailed studies of the underlying astrophysical processes at work in the source. Note, however, that stellar-mass BBHs are only seconds away from merging when they leave the B-DECIGO frequency range at $10\,\text{Hz}$, so to determine the sky localization before the merger takes place, one is limited to the information accumulated in B-DECIGO until a few minutes or hours before the merger takes place. Still, even with somewhat reduced information the estimates of the sky localization will be very precise.




\subsection{Average parameter estimates}

\begin{table*}
\begin{tabular}{c|c|c|c|c|c|c|c|c}
• & HLV & ET & BDEC  & BDEC2 & HLV/BDEC & ET/BDEC & HLV/ET  & HLV/ET/BDEC \\ 
\hline 
$\Delta \mu$ / $ \mu$ & $1.9 \times 10^{-2}$  & $3.8 \times 10^{-4}$ & $1.9 \times 10^{-4}$ & $2.2 \times 10^{-4}$ & $1.2 \times 10^{-5}$  & $2.5 \times 10^{-5}$ & $3.5 \times 10^{-4}$ & $2.5 \times 10^{-5}$ \\ 
\hline 
$\Delta \Mtot$ / $\Mtot$ & $2.5 \times 10^{-2}$  & $5.2 \times 10^{-4}$ & $2.8 \times 10^{-4}$ & $3.2 \times 10^{-4}$ & $1.7 \times 10^{-4}$  & $3.8 \times 10^{-5}$ & $4.9 \times 10^{-4}$ & $3.8 \times 10^{-5}$ \\ 
\hline 
$\Delta r$ / $r$ & $1.44$ & $1.38$ & $0.14$ & $0.26$ & $0.14$ & $0.07$ & $0.3$ & $0.07$ \\ 
\hline 
$\Delta \cos(\iota) $ & $1.37$ & $1.2$ & $0.15$ & $0.24$ & $0.13$ & $0.07$ & $0.28$ & $0.07$ \\ 
\hline 
$\Delta \theta / \deg$ & $10.3$ & $62$ & $0.31$ & $0.63$ & $0.16$ & $0.13$ & $2.1$ & $0.12$ \\ 
\hline 
$\Delta \phi / \deg$ & $12.1$ & $31.1$ & $0.23$ & $0.44$ & $0.1$ & $0.08$ & $2.65$ & $0.08$ \\ 
\hline 
$\Delta \psi / \deg$ & $210$ & $198$ & $49$ & $64$ & $48$ & $33$ & $70$ & $33$ \\
\hline 
$\Omega / \deg^2$ & $277$ & $7207$ & $0.18$ & $0.52$ & $5 \times 10^{-4}$ & $2.3 \times 10^{-4}$ & $13$ & $2.2 \times 10^{-4}$ \\
\end{tabular}
\caption{Fisher estimates for the standard deviations, averaged over 100 stellar-mass BBHs that are detected in all detectors. BDEC is B-DECIGO with a geostationary orbit, BDEC2 with a LISA-like orbit. Note that the errors on the mass parameters are expressed relative to the mass parameters (a difference with respect to table~\ref{table:errors}).}
\label{table:averageerrors}
\end{table*}

\begin{figure*}[!ht]
   \centering
   	(a)\includegraphics[width=.45\textwidth]{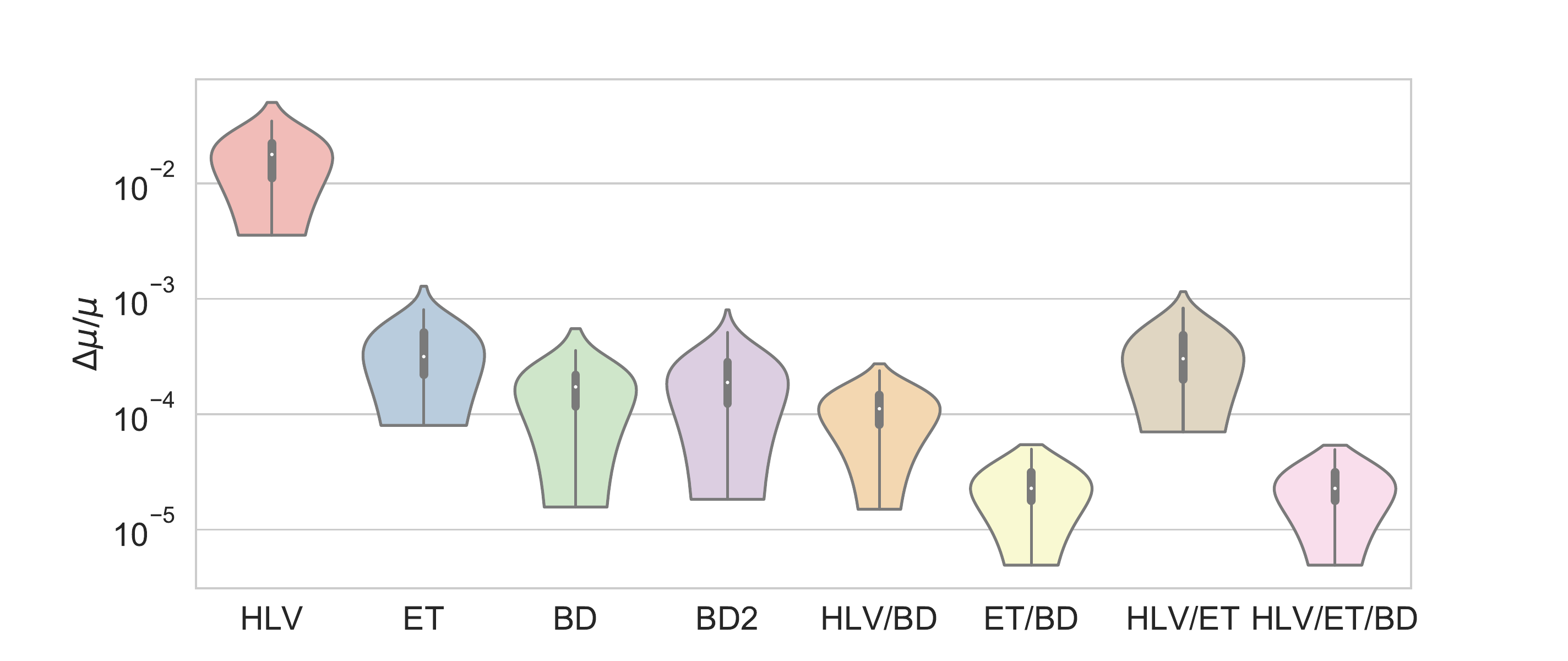}
   	(b)\includegraphics[width=.45\textwidth]{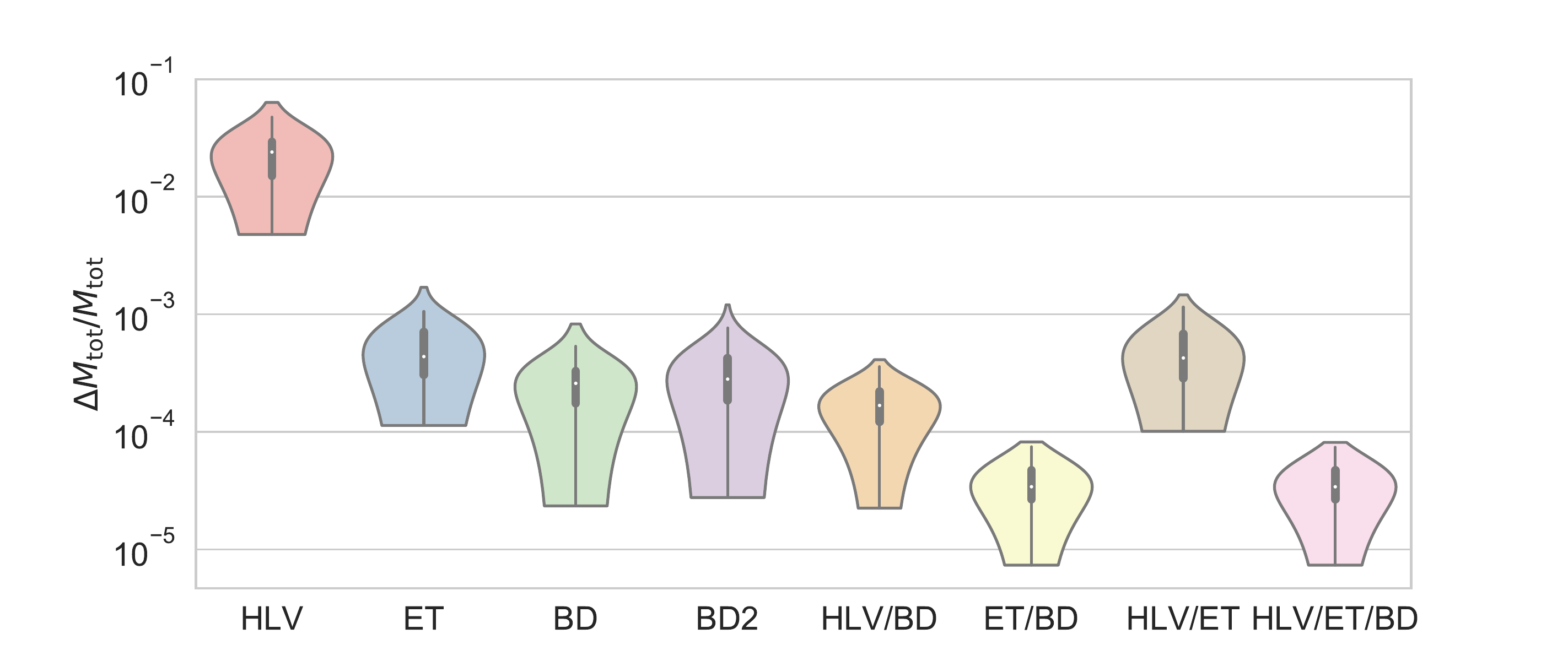}\\
	(c)\includegraphics[width=.45\textwidth]{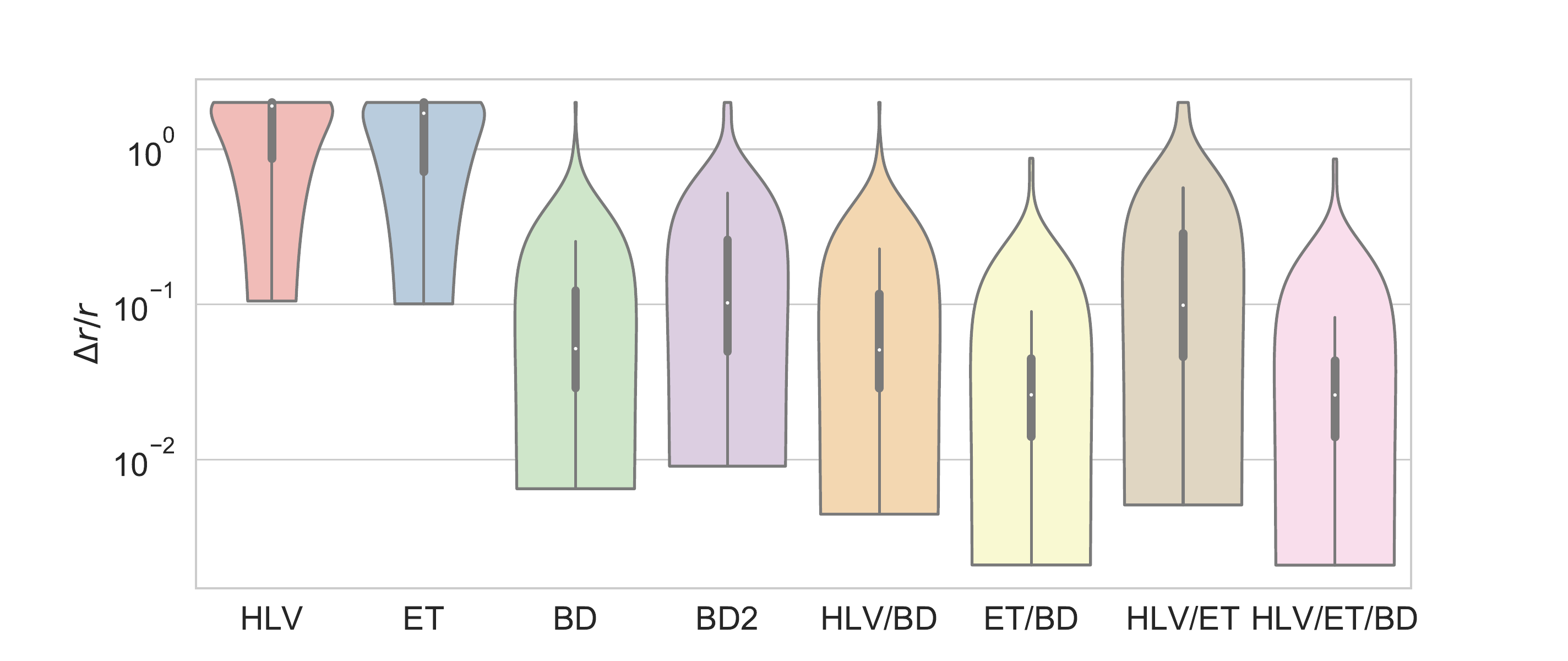}
	(d)\includegraphics[width=.45\textwidth]{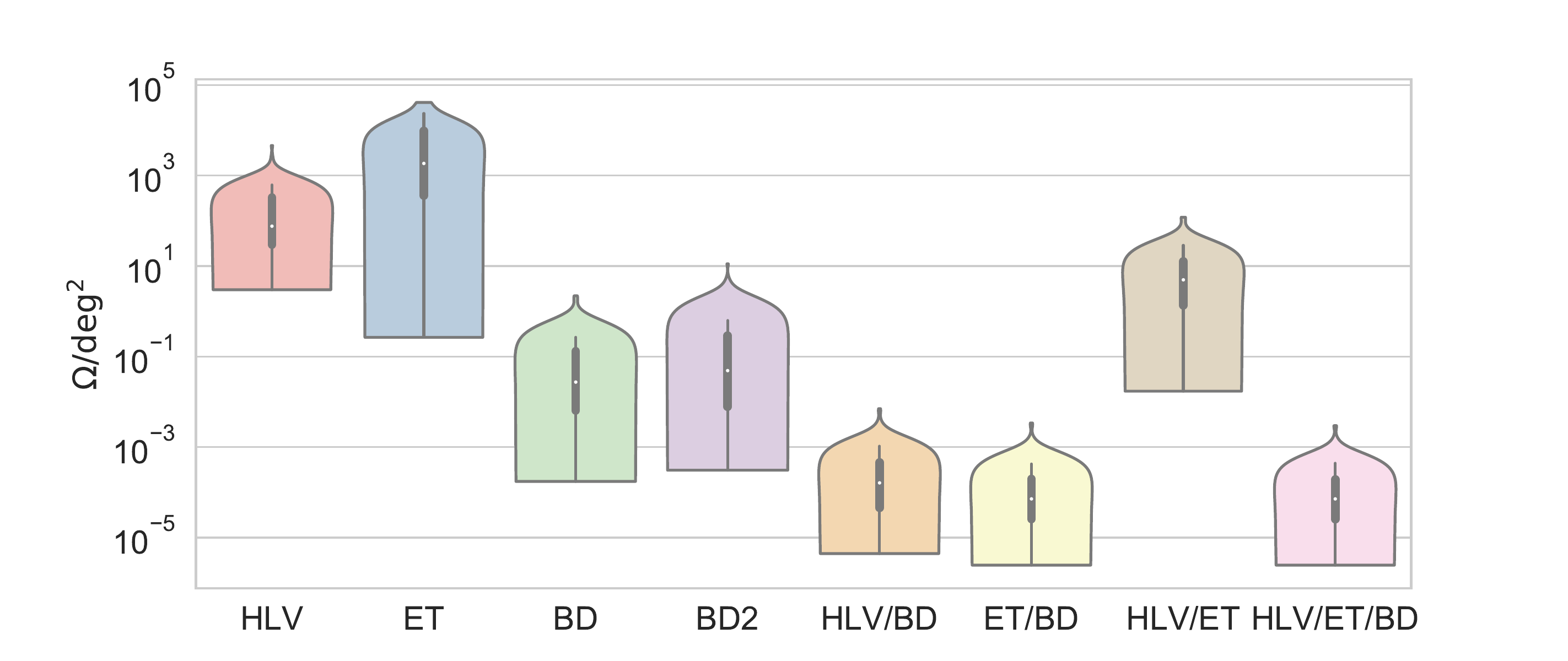}\\
	(e)\includegraphics[width=.45\textwidth]{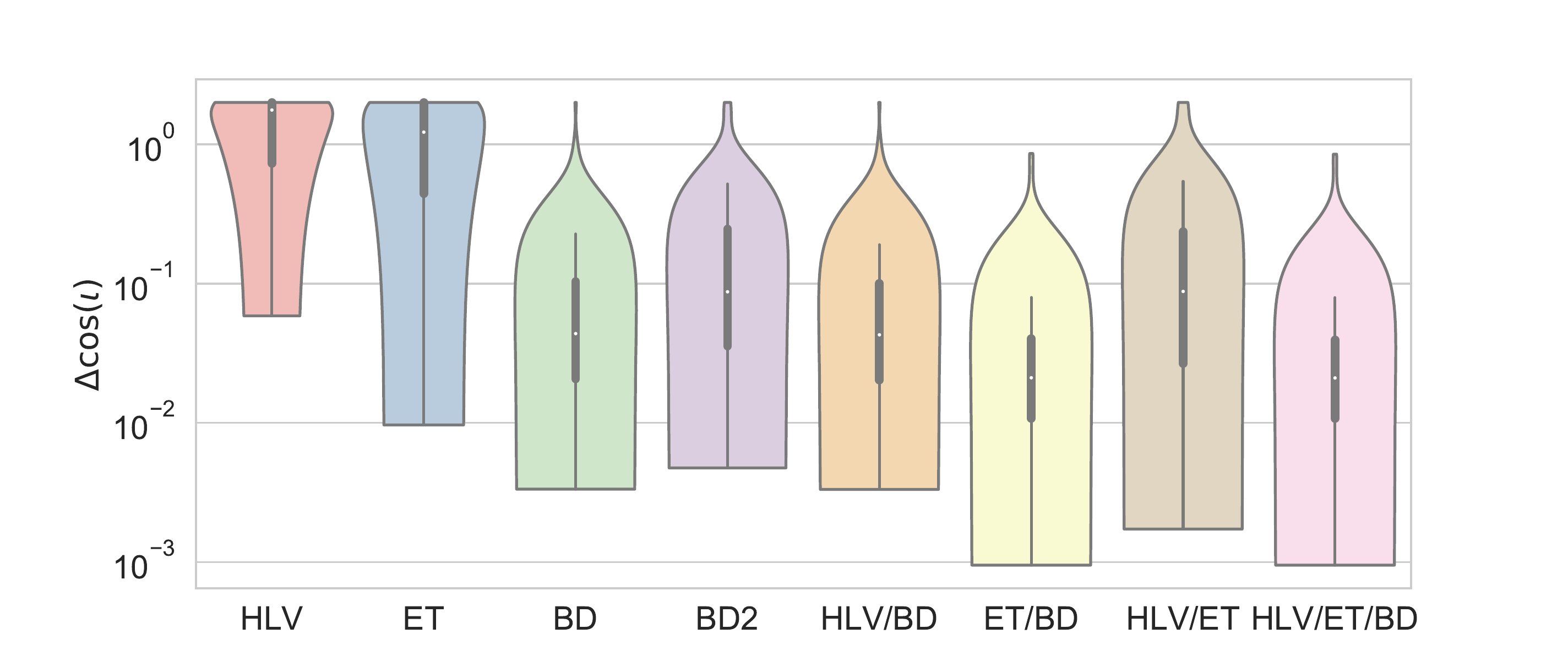}
	(f)\includegraphics[width=.45\textwidth]{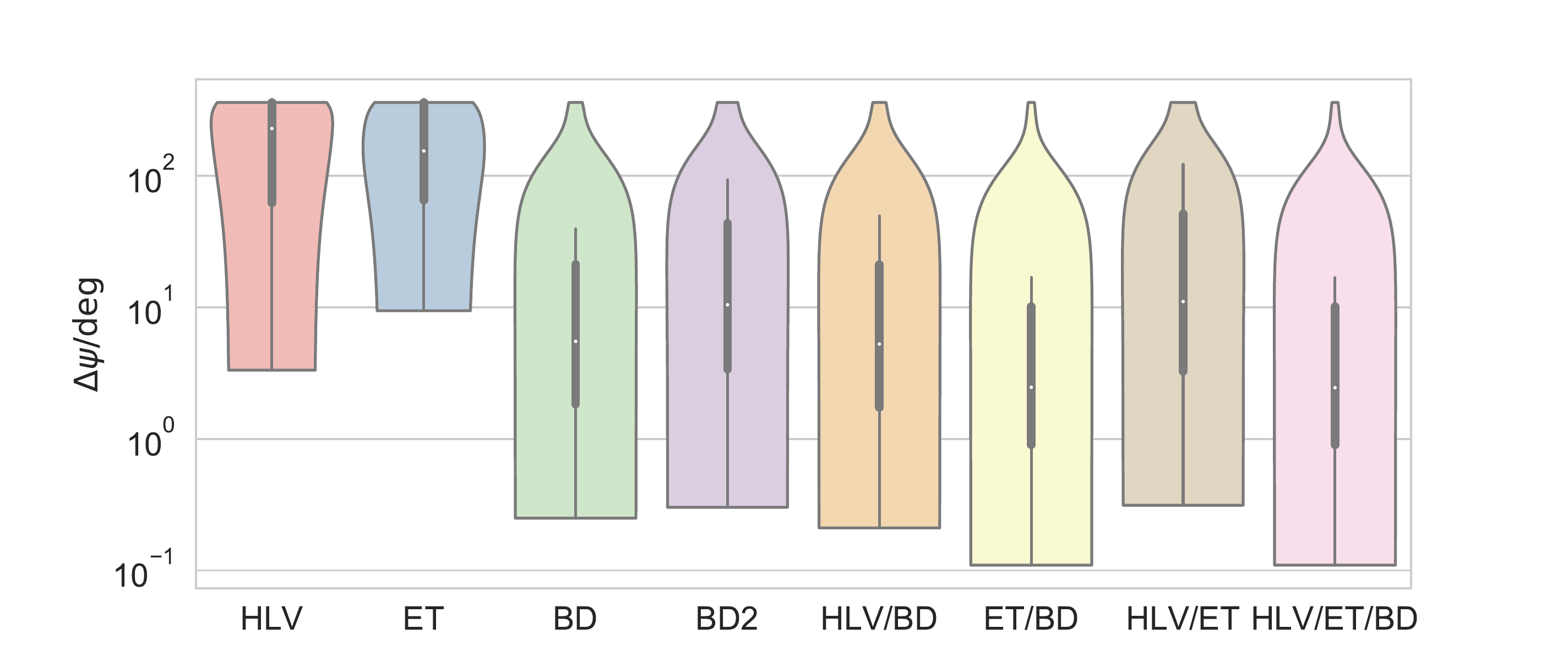}\\
   \caption{Distributions of the standard deviations of the estimates of the stellar-mass BBHs, for the mass parameters ((a) + (b)), the distance (c), the sky localization (d), the inclination angle (e), and the polarization angle (f). In this plot, BD, BD2 stand for BDEC, BDEC2.}
   \label{fig:ViolinStellar}
\end{figure*}

Now, we take a look at the results we obtain when we average the parameter estimates of many binaries. Table~\ref{table:averageerrors} shows the standard deviations we obtain after averaging over 100 events detected in all the detectors. Note that here, we consider the standard deviations of the mass parameters relative to their true values (a difference to table~\ref{table:errors}). Similarly, in this study, we limited the maximum standard deviation for the angles to $360\,\deg$. In the case of the inclination angle $\iota$, we show the standard deviation on $\cos(\iota)$, and limit the maximum error to $2$. This is necessary to prevent single events with a large standard deviation of a Gaussian distribution, which poorly approximates the actual distributions in these cases, from dominating the overall estimate. We also restricted the estimate of the relative standard deviation of the distance to 2 (the maximum uncertainty of the distance is 2x the distance itself). It is also important to emphasize that the averages involving the more sensitive detectors ET and B-DECIGO include only events that are also detected in HLV, and thus have a higher SNR and better parameter estimates than we would obtain from an average over all the binaries detected in these detectors (the selection of events that are also detected in HLV represents a bias towards high-SNR events when looking at all the events detectable in ET or B-DECIGO).

In general, the average parameter estimates for the cosmological distribution are worse than the estimates for the example binary. This is due to the fact that the example binary is relatively close, and is detected with a high SNR. The SNR distribution shown in fig.~\ref{fig:SNRdist1} shows that the vast majority of the binaries in the cosmological distribution is detected with a lower SNR, and will therefore have larger standard deviations of the parameter estimate.

The averaged mass estimates are one to two orders of magnitude larger than the ones from the example binary.  The same is true for the sky localization. We can see that unlike in the example binary, in the averaged mass parameter estimates BDEC gives better mass parameter estimates than ET. Unlike in the example binary, distance, polarization angle and inclination angle remain essentially undetermined in all detectors except BDEC. Apparently, detections of binaries with lower SNR do not provide sufficient information to determine these parameters. Only BDEC is able to provide some information on the distance in particular, and the differences between BDEC and BDEC2 are significantly larger than for the example binary -- for events with lower SNR, the differences in detector movement play a larger role. The sky localization in BDEC is significantly better than in BDEC2. With respect to the average sky localization of B-DECIGO (BDEC), it is worth noting that it is good enough to allow for multi-messenger astronomy.

In fig.~\ref{fig:ViolinStellar}, we plot the distributions of the parameter estimates of the mass parameters, the distance, the sky localization, the cosine of the inclination angle, and the polarization angle. A salient feature of the distributions of ET is the fact that the distributions of the sky localization, the distance, and the inclination and polarization angles are very wide. While for the majority of events ET can provide little information about these parameters, there are a few events for which ET can provide very precise estimates, with sky localizations below $10\,\deg^2$ and an error on the distance of $10 \%$. It seems plausible that it is the triangular shape of ET that allows the detector to determine these parameters so precisely for a few selected, high-SNR events, despite the fact that ET neither benefits from time triangulation nor from detector movement effects.

Summing up the results, we have seen that B-DECIGO would benefit from having a geostationary orbit, rather than a LISA-like orbit, due to the advantages of a shorter time scale of the detector movement. B-DECIGO could determine the source position very precisely, and before the merger takes place, which would allow for the capture of a possible electromagnetic counterpart. In addition to that, it would significantly improve the estimate of the mass parameters, and it is the only detector that can disentangle the degeneracy between the parameters $\theta$, $\phi$, $\psi$, $\iota$, and $r$ effectively. With respect to multiband parameter estimation, the mass estimate benefits greatly from combining B-DECIGO with ground-based detectors (HLV or ET). ET will also significantly improve the estimate of the mass parameters. In combination with the HLV network, it will also improve the sky localization of the source via time triangulation, but unlike B-DECIGO, it cannot determine the source position before the merger takes place. Also, in combination with the HLV network, ET can improve the estimate of the inclination angle $\iota$ and the polarization angle $\psi$.

\section{Intermediate-mass Binary Black Holes (IMBHs): ET, B-DECIGO, and LISA}
\label{sec:IMBH}

In this section, we investigate how well we can estimate the parameters of intermediate-mass black hole (IMBH) binaries with the different detector networks. A study of the multiband observations of IMBH inspiral phases was recently published \cite{JaEA2019}. The black-hole binaries we are looking at have masses in the range of a few hundred solar masses. These binaries merge in the frequency range of 1--20\,Hz, which borders on the lower end of the sensitivity band of the HLV detectors. Consequently, the currently operating detectors will only be able to see mergers of the lightest IMBHs, and not very well. Therefore, we only include ET, B-DECIGO, and LISA in this analysis. Due to their higher masses, IMBH binaries can be observed in LISA with sufficient SNR.

\subsection{Detection capabilities of the detectors}

\begin{figure}[t]
\centering
\includegraphics[scale=0.5]{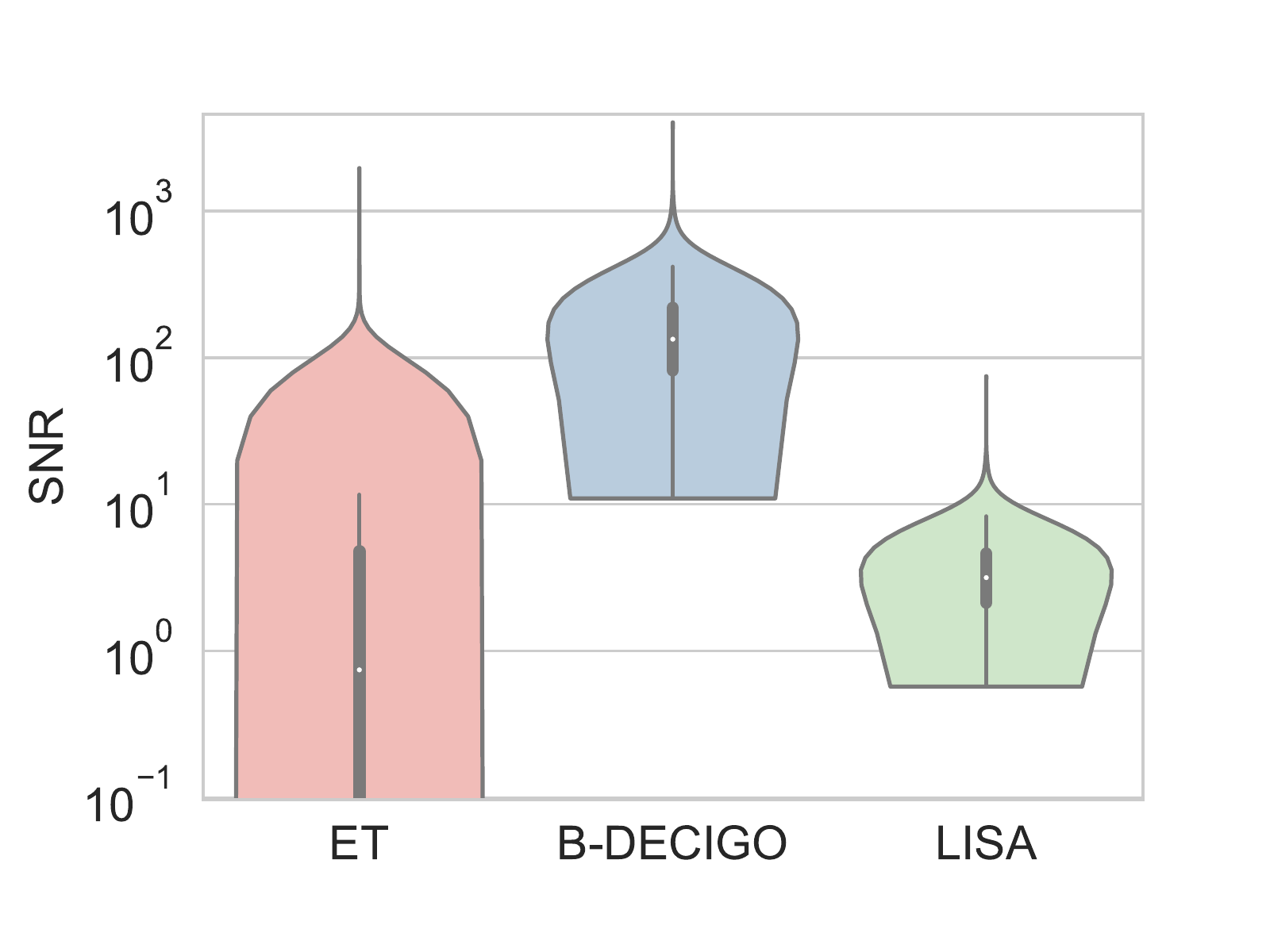}
\caption{Distribution of SNRs in ET, B-DECIGO, and LISA, for a cosmological distribution of 1000 intermediate-mass binary black holes.}
\label{fig:SNRdistIMBH}
\end{figure}

Like in the previous section, we first perform a population study of 1000 IMBH binaries to investigate the detection capabilities of the detectors. As mentioned before, it is not clear whether IMBHs exist or not, and if they do, what their mass distribution is. Therefore, we use the same population model as in the previous section (redshift-independent power-law with index $1.6$, spatial distribution as in fig.~\ref{fig:Pz}), as a toy model. In the case of stellar-mass BBHs, the main justifications for this assumption are that a power law is a simple model and that, since stellar-mass BHs are thought to originate from stars, then it is reasonable to assume that their mass function follows a power law as that of stars does \cite{Ses2016}. This is not necessarily true for IMBHs, whose formation mechanism can be different \cite{mapelli}. A recent study suggests that IMBHs of masses up to $\approx 440 \msolar$ can form via merging processes in young star clusters \cite{DCEA2019}, but the number of the IMBHs simulated is too small to draw a conclusion. Therefore, we want to mention at this point that there is \textit{no physical motivation for choosing the model we use}, and that we simply use it for lack of a better alternative. We choose the mass range of the individual IMBHs to be $150 \msolar<m_\text{IMBH}<500 \msolar$. The distributions of the SNR are given in fig.~\ref{fig:SNRdistIMBH}.
\begin{table}
\begin{tabular}{c|c|c|c}
$\text{SNR}>10$& ET & B-DECIGO & LISA \\ 
\hline 
out of 1000 events & 128 & 1000 & 28 \\ 
\end{tabular}
\caption{Number of intermediate-mass binary black hole events with $\text{SNR}>10$, in different detectors.}
\label{table:events}
\end{table}
As before, for the Fisher-matrix approach, we require a minimum SNR of 10. The number of events that meet this requirement is shown in table~\ref{table:events}. We can see that B-DECIGO is by far the best detector to detect IMBHs. It detects all the 1000 IMBH binaries. Due to the lower sensitivity, the SNRs in LISA are lower than in B-DECIGO, but still LISA is able to detect 28 out of 1000 IMBH binaries. Despite its high sensitivity, ET is only able to detect 128 IMBH binaries. This is due to the fact that only light and relatively close IMBH binaries are detectable in ET, the heavier and more distant ones merge at frequencies that are too low to be detected in ET (distance enters because the masses are redshifted, such that more distant sources appear to be heavier in the detector frame). Higher Harmonics, which we neglected in this study, can contribute significantly to the SNR of IMBHs in ET \cite{BrSe2006}. Including them could increase the mass and distance range of IMBHs detectable with ET, so it is important to note at this point that the result we obtain may overestimate standard deviations of the parameter estimates in ET and underestimate SNRs. 

When looking at the SNR-distribution in ET, we see that it spreads over a large range - some binaries are detected with a very high SNR, but many only with an extremely low SNR. The ability of ET to detect IMBH binaries is sensitive to the mass distribution - a mass distribution that favors lighter IMBHs would yield a higher number of events detected in ET. This is exactly opposed to the space-based detectors, whose detection capability benefits from heavier IMBHs. Already at this point, we can make a statement about the scientific value of B-DECIGO: it would doubtlessly answer the question whether IMBHs exist, and if they do exist, it could tell us how many there are, and what their mass spectrum is.

\subsection{An example event}

Now we take a look at an example event. The event we are looking at is the merger of a binary with a black hole of mass $m_1 = 200 \msolar$ and another black hole of mass $m_2 = 150 \msolar$, at a distance of $r = 1000 \Mpc$. The SNR of this event in the detectors is shown in table~\ref{table:SNRIMBH}. The estimates for the standard deviations is shown in table~\ref{table:errorsIMBH}.

\begin{table}
\begin{tabular}{c|c|c|c}
• & ET & BDEC & LISA \\ 
\hline 
SNR & 311 & 872 & 15 \\ 
\end{tabular}
\caption{SNR of the example binary, in the detectors we investigate.}
\label{table:SNRIMBH}
\end{table}

\begin{table*}
\begin{tabular}{c|c|c|c|c|c|c|c|c}
• & ET & BDEC & BDEC2 & LISA & ET/LISA & ET/BDEC & BDEC/LISA & ET/BDEC/LISA \\ 
\hline 
$\Delta \mu$ / $\msolar$ & $4.7 \times 10^{-1}$  & $5.5 \times 10^{-3}$ & $5.1 \times 10^{-3}$ & $9.5 \times 10^{-1}$ & $2.1 \times 10^{-2}$ & $4.8 \times 10^{-3}$ & $1.8 \times 10^{-3}$ & $1.7 \times 10^{-4}$ \\ 
\hline 
$\Delta \Mtot$ / $\msolar$ & $2.3$  & $3.3 \times 10^{-2}$ & $3 \times 10^{-2}$ & $5.8$ & $1.3 \times 10^{-1}$ & $2.9 \times 10^{-2}$ & $1.1 \times 10^{-2}$ & $1.1 \times 10^{-3}$ \\ 
\hline 
$\Delta r$ / $\Mpc$ & $26306$  & $7.3$ & $63$ & $423$ & $17$ & $6.5$ & $6.7$ & $6.1$  \\ 
\hline 
$\Delta \iota / \deg$ & $360$  & $0.48$ & $6.7$ & $28$ & $1.24$ & $0.43$ & $0.43$ & $0.4$ \\ 
\hline 
$\Delta \theta / \deg$ & $360$  & $0.11$ & $1.8$ & $0.44$ & $0.36$ & $0.1$ & $0.1$ & $0.1$  \\ 
\hline 
$\Delta \phi / \deg$ & $109$  & $0.013$ & $0.01$ & $0.3$ & $0.18$ & $1 \times 10^{-4}$ & $0.013$ & $1 \times 10^{-4}$ \\ 
\hline 
$\Delta \psi / \deg$ & $360$  & $0.71$ & $17.9$ & $35.9$ & $1.7$ & $0.64$ & $0.69$ & $0.63$ \\
\hline 
$\Omega / \deg^2$ & $41253$  & $4.6 \times 10^{-3}$ & $5.9 \times 10^{-2}$ & $0.41$ & $0.2$ & $3.6 \times 10^{-5}$ & $4.2 \times 10^{-3}$ & $3.1 \times 10^{-5}$  \\
\end{tabular}
\caption{Fisher estimates for the standard deviations of the example IMBH binary, for different detectors and detector networks. BDEC is B-DECIGO with a geostationary orbit, BDEC2 with a LISA-like orbit.}
\label{table:errorsIMBH}
\end{table*}

\begin{figure}[t]
\centering
\includegraphics[scale=0.5]{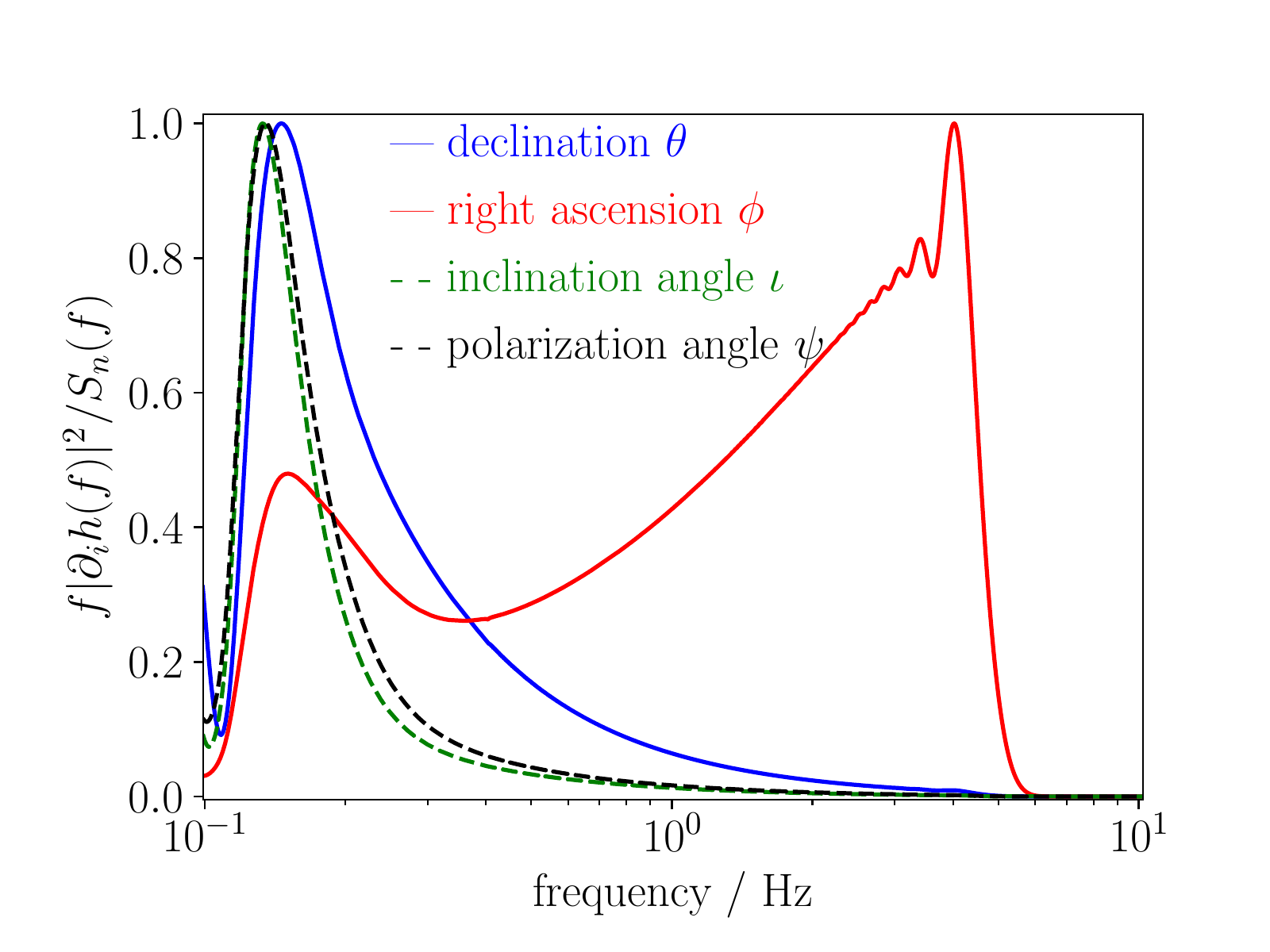}
\caption{Frequency-dependence of the information content of the diagonal element of the Fisher-matrix in BDEC, for the example IMBH binary.}
\label{fig:IMBHBDECinfoplot}
\end{figure}

\begin{figure}[t]
\centering
\includegraphics[scale=0.5]{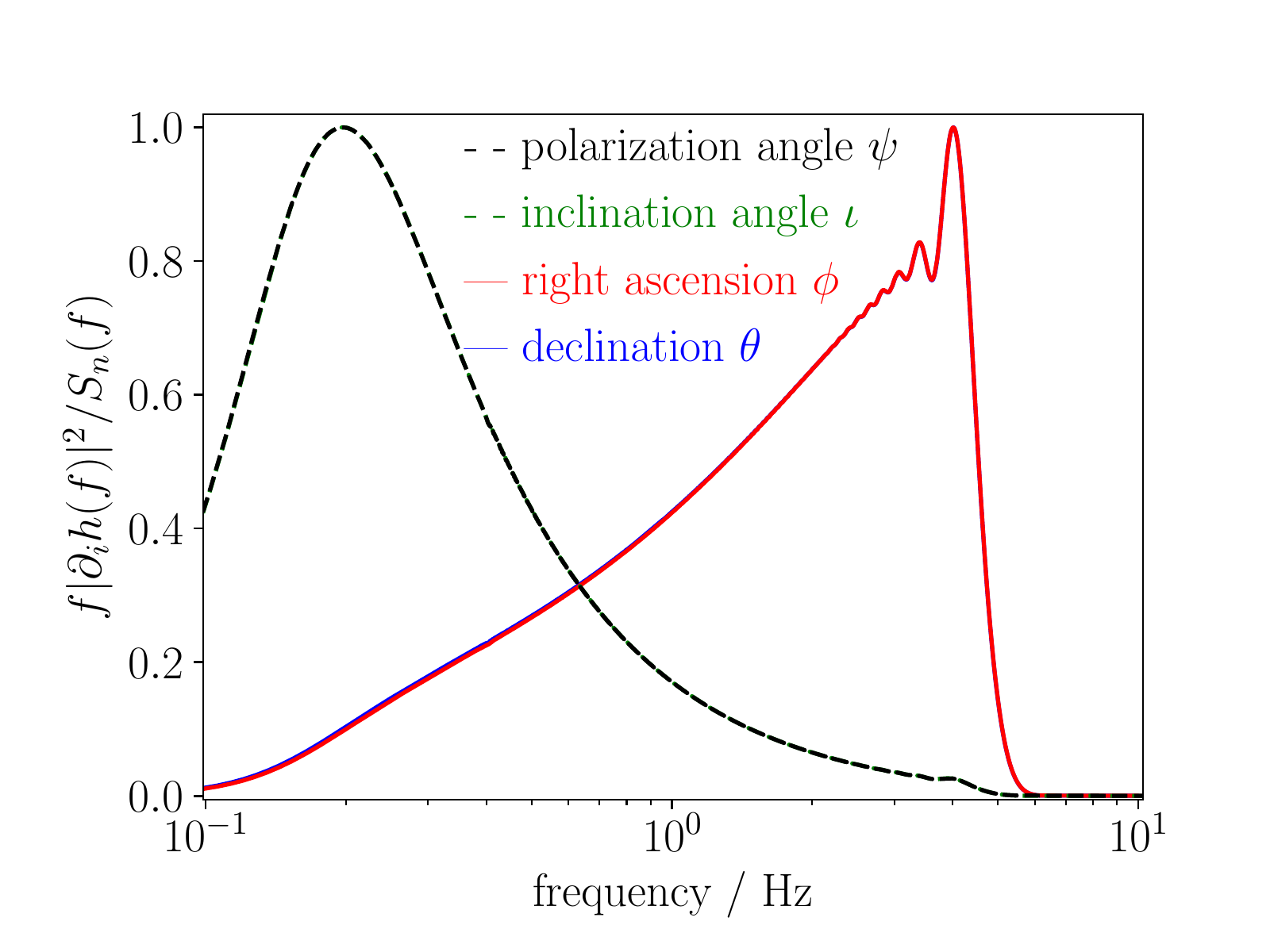}
\caption{Frequency-dependence of the information content of the diagonal element of the Fisher-matrix in BDEC2, for the example IMBH binary.}
\label{fig:IMBHBDEC2infoplot}
\end{figure}

We can see in table~\ref{table:SNRIMBH} that for this binary, which is rather close, ET and B-DECIGO yield extremely high SNRs. LISA is significantly less sensitive (again, assuming a 3-year observation time), but the SNR is still sufficiently high for a Fisher analysis.
With respect to the parameter estimation, the first thing we notice is the difference between the geostationary orbit and the LISA-like orbit in B-DECIGO (BDEC and BDEC2). The effects we described in the previous section for the stellar-mass BBHs is even more pronounced for IMBHs, because these are heavier and spend less time in the B-DECIGO frequency range. Therefore, the estimate of sky localization, inclination and polarization angles and distance are significantly worse in BDEC2. To illustrate the influence of movement effects, we again plot the function $f|\partial_i h(f)|^2/S_n(f)$ for BDEC and BDEC2, in fig.~\ref{fig:IMBHBDECinfoplot} and fig.~\ref{fig:IMBHBDEC2infoplot}. We see that in BDEC2 (fig.~\ref{fig:IMBHBDEC2infoplot}), the sky position angles are perfectly correlated - the observation time is so short that movement effects play no role at all. The same is true for the inclination and polarization angles. In BDEC, the movement effects help to break the degeneracies.

BDEC gives the most precise parameter estimates for all parameters. LISA can also determine sky localization, inclination and polarization angles and distance, but with standard deviations orders of magnitude larger than in BDEC. In ET, these parameters are entirely undetermined. B-DECIGO gives much better estimates of the mass parameters than LISA. This is due to the effect already mentioned before: far away from merger, the mass parameters are degenerate, which limits the estimate of the individual mass parameters. LISA can measure the chirp mass very precisely, but for a precise estimate of the individual mass parameters it needs information from a detector that can detect the gravitational wave close to merger. For this reason, the estimate of the mass parameters of the combination ET/LISA is significantly better than the estimate of the individual detectors. The estimate of the combination BDEC/LISA is approximately a factor of 2--3 better than BDEC alone. Since most IMBHs merge in the frequency range of 1--20\,Hz, B-DECIGO can also detect the waveform close to merger, and there is no problem with the degeneracy of the mass parameters in B-DECIGO. The estimates of the inclination and polarization angles is dominated by BDEC - adding information from other detectors improves the estimates only slightly. Combining ET and BDEC improves the estimate of the sky localization.

It is important to note that the ability of ET to reconstruct the properties of the binary depend very strongly on the mass. Heavier binaries cannot be detected in ET, since they merge at too low frequencies. Conversely, lighter binaries spend more time in the ET band and will provide ET with more information. As mentioned before, the simplifications of our model may significantly affect the results, and the inclusion of higher harmonics could make a large difference for some signals.

With respect to the determination of the sky localization in B-DECIGO, it is important to note that for most IMBH binaries, the merger takes place in the B-DECIGO frequency range, and the observation times are less than a day. This means that the full sky localization information is available only after the merger takes place. If one wishes to determine the sky localization before the merger takes place, one can do so by using the reduced information available from the detection up to a few minutes before the merger. LISA is sensitive at lower frequencies and does not detect the merger waveform, so the LISA estimate is always available before merger.

\subsection{Average parameter estimates}

\begin{table*}
\begin{tabular}{c|c|c|c|c|c|c|c|c}
• & ET & BDEC & BDEC2 & LISA & ET/LISA & ET/BDEC & BDEC/LISA & ET/BDEC/LISA \\ 
\hline 
$\Delta \mu$ / $\mu$ & $2.8 \times 10^{-2}$  & $6.8 \times 10^{-5}$ & $6.6 \times 10^{-5}$ & $6.1 \times 10^{-3}$ & $3.8 \times 10^{-4}$ & $6.7 \times 10^{-5}$ & $2.4 \times 10^{-5}$ & $2.3 \times 10^{-5}$ \\ 
\hline 
$\Delta \Mtot$ / $\Mtot$ & $3.7 \times 10^{-2}$  & $9.9 \times 10^{-5}$ & $9.6 \times 10^{-5}$ & $9.1 \times 10^{-3}$ & $5.7 \times 10^{-4}$ & $9.7 \times 10^{-5}$ & $3.6 \times 10^{-5}$ & $3.4 \times 10^{-5}$ \\ 
\hline 
$\Delta r$ / $r$ & $2$  & $0.056$ & $0.62$ & $1.02$ & $0.48$ & $0.047$ & $0.053$ & $0.046$  \\ 
\hline 
$\Delta \cos(\iota)$ & $1.97$  & $0.054$ & $0.58$ & $0.97$ & $0.46$ & $0.045$ & $0.05$ & $0.04$ \\ 
\hline 
$\Delta \theta / \deg$ & $352$  & $0.2$ & $4.9$ & $0.98$ & $0.63$ & $0.046$ & $0.16$ & $0.044$  \\ 
\hline 
$\Delta \phi / \deg$ & $335$  & $0.034$ & $0.056$ & $0.61$ & $0.4$ & $0.032$ & $0.032$ & $0.03$ \\ 
\hline 
$\Delta \psi / \deg$ & $358$  & $37$ & $123$ & $177$ & $104$ & $33$ & $35$ & $33$ \\
\hline 
$\Omega / \deg^2$ & $41248$  & $2.6 \times 10^{-2}$ & $1.7$ & $1.3$ & $0.51$ & $8.8 \times 10^{-5}$ & $1.9 \times 10^{-2}$ & $8 \times 10^{-5}$  \\
\end{tabular}
\caption{Fisher estimates for the standard deviations, averaged over 100 IMBH binaries that are detected in all detectors. BDEC is B-DECIGO with a geostationary orbit, BDEC2 with a LISA-like orbit. Note that the errors on the mass parameters are expressed relative to the mass parameters (a difference with respect to table~\ref{table:errorsIMBH}).}
\label{table:avergageerrorsIMBH}
\end{table*}

\begin{figure*}[!ht]
   \centering
   (a)\includegraphics[width=.45\textwidth]{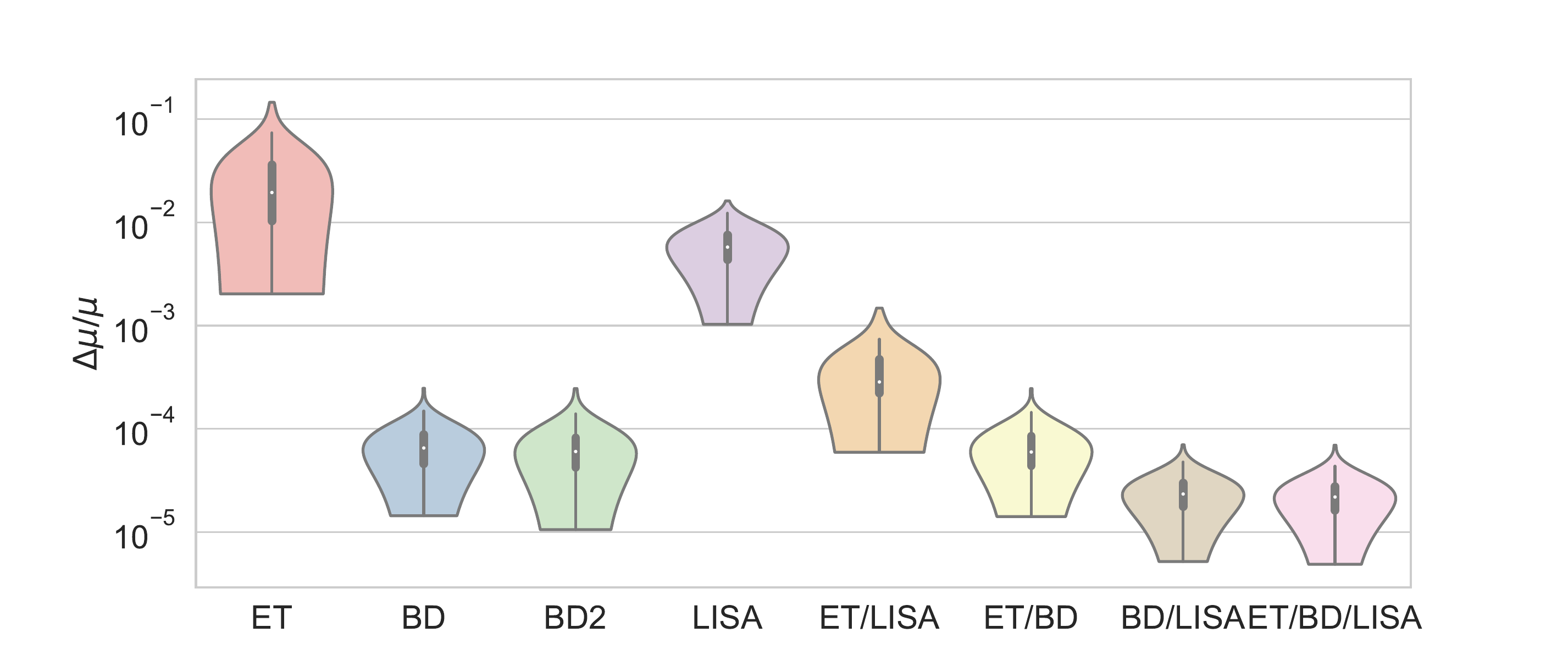}
   (b)\includegraphics[width=.45\textwidth]{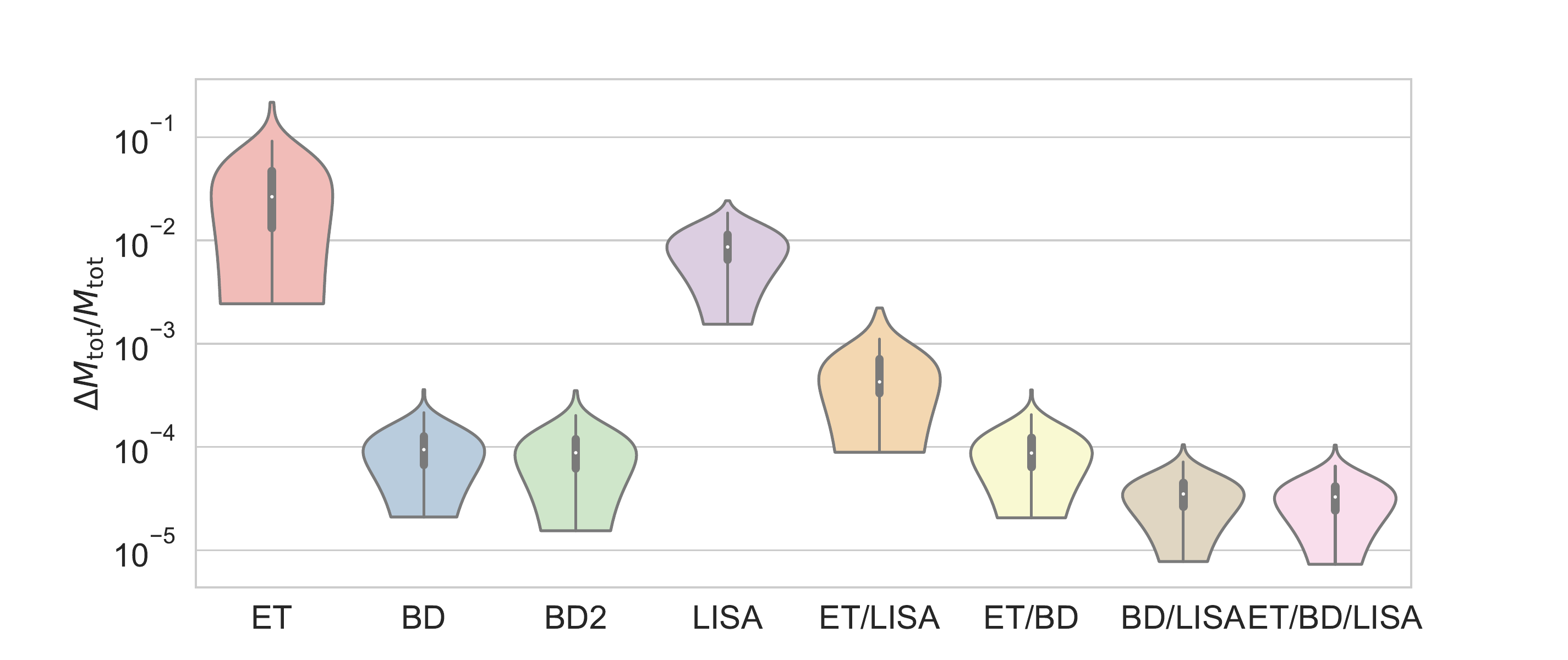}\\
   (c)\includegraphics[width=.45\textwidth]{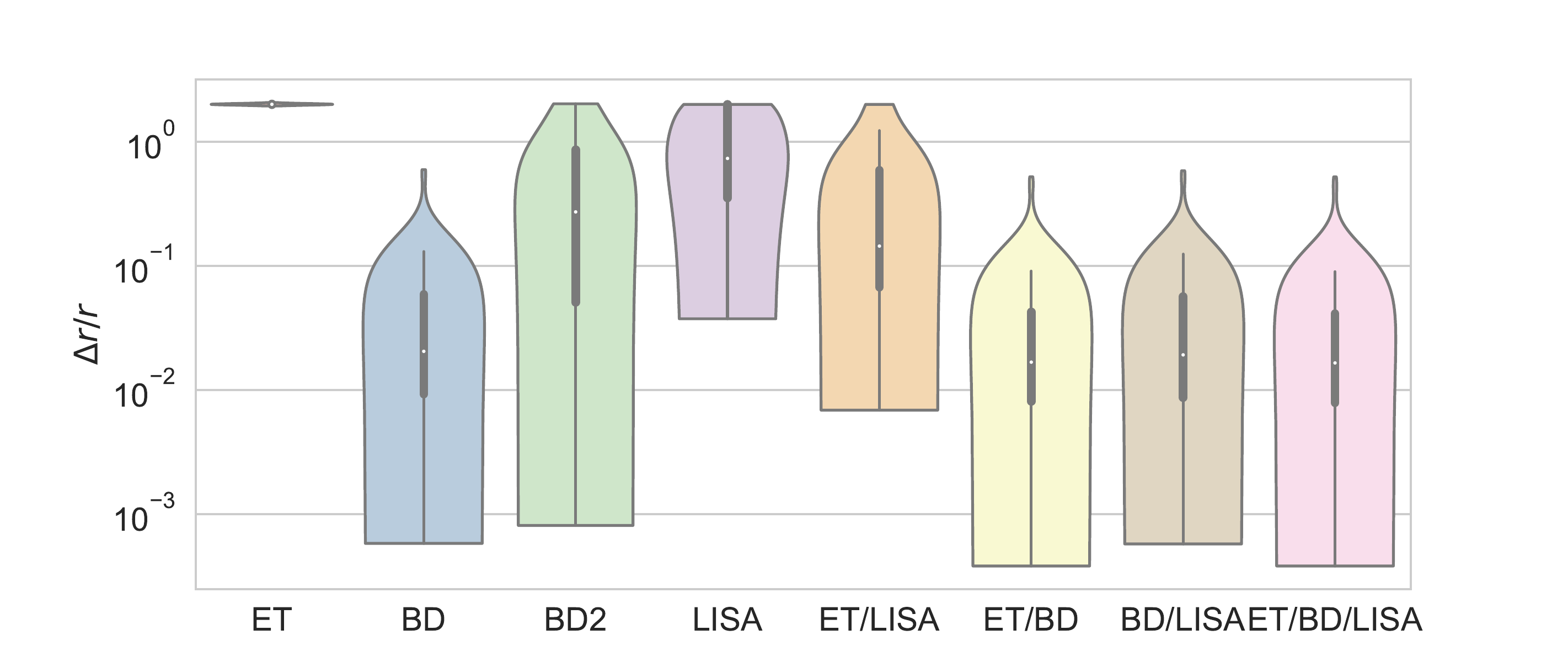}
   (d)\includegraphics[width=.45\textwidth]{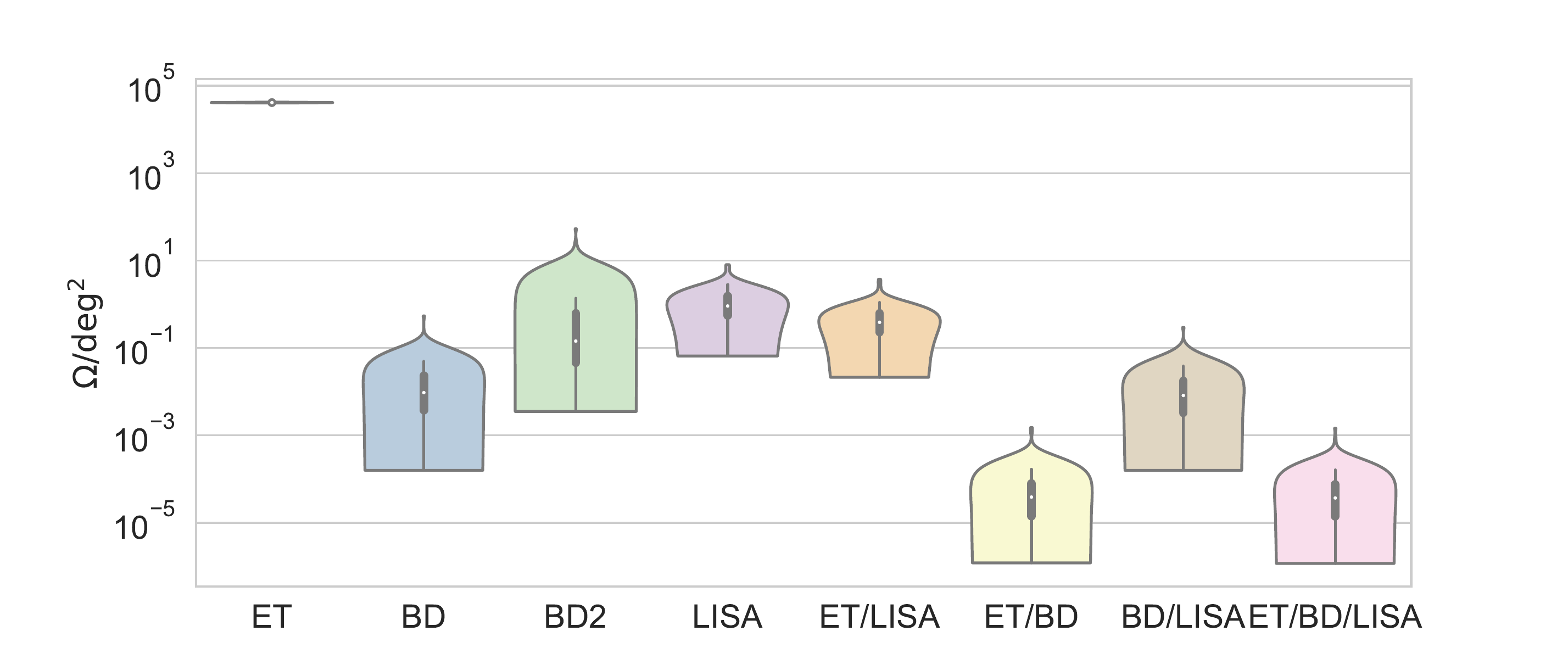}\\
   (e)\includegraphics[width=.45\textwidth]{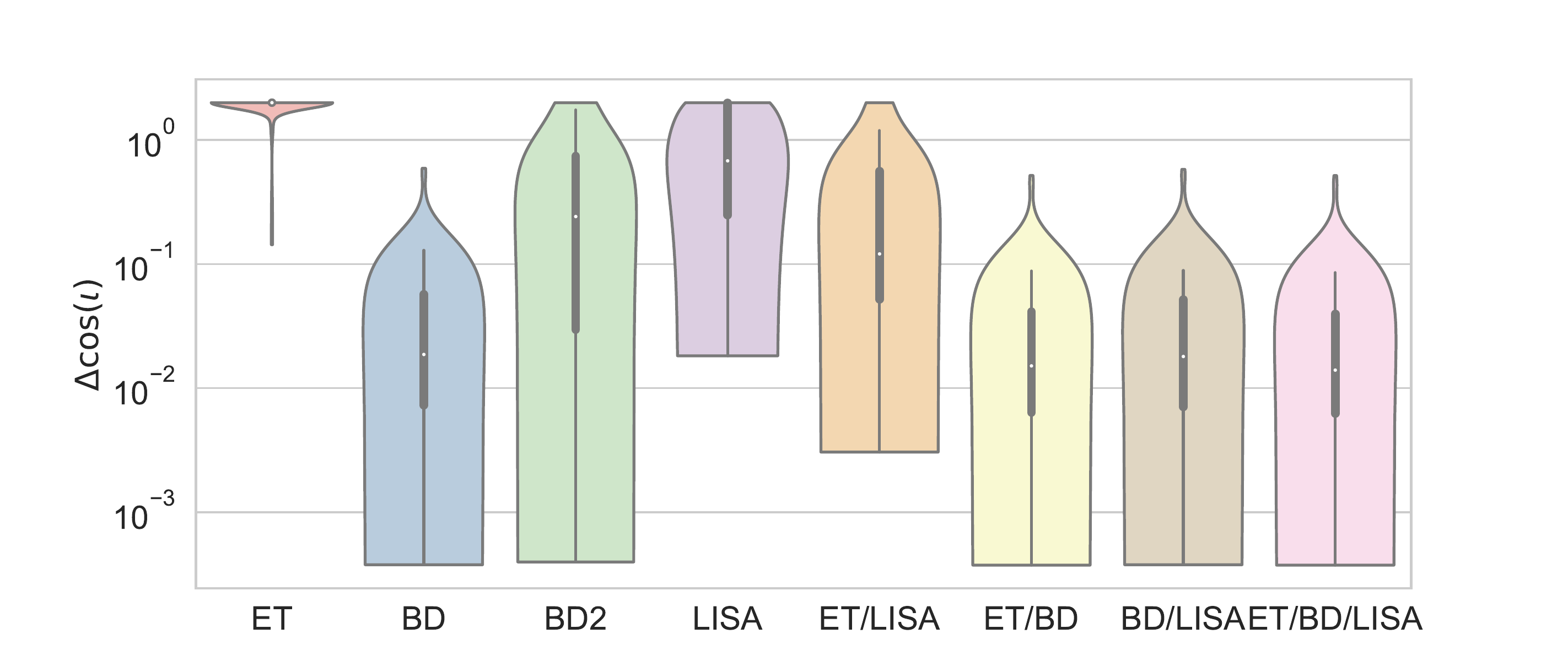}
   (f)\includegraphics[width=.45\textwidth]{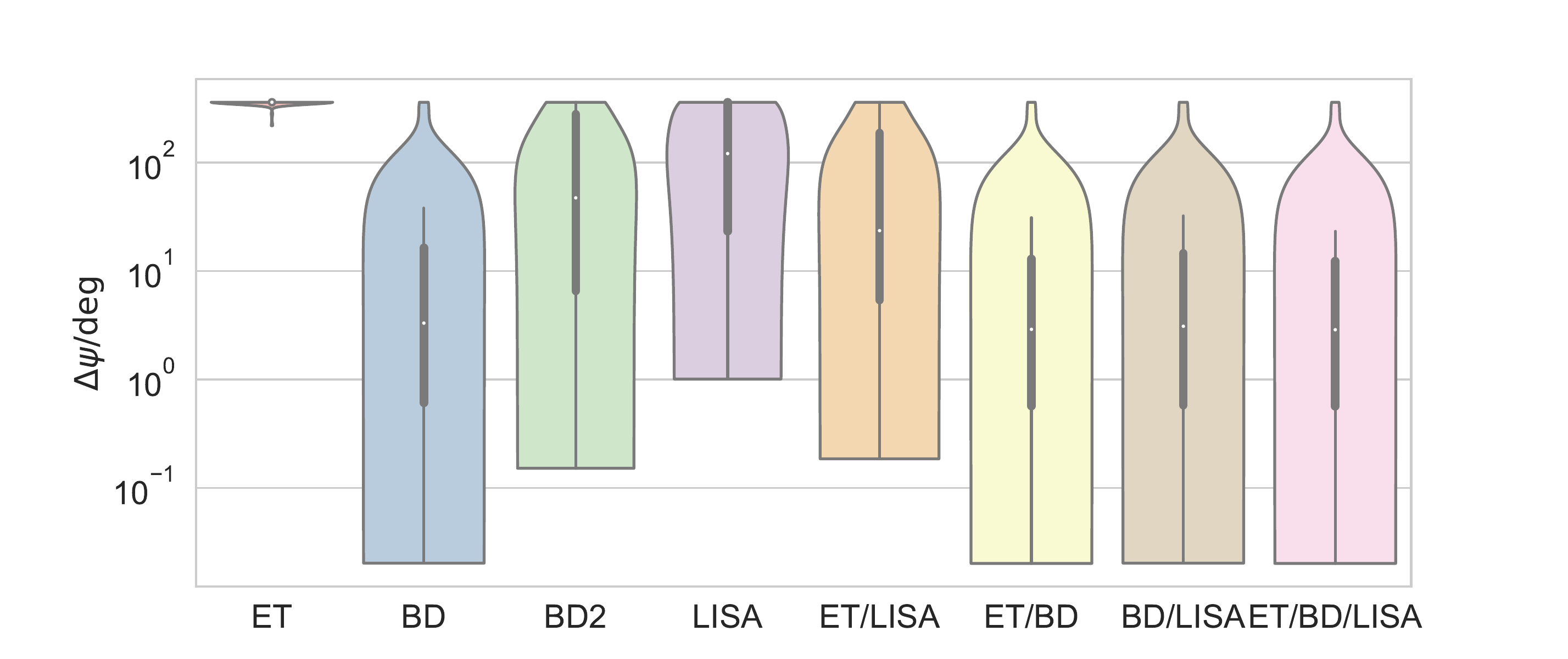}\\
   \caption{Distributions of the standard deviations of the parameter estimate of the IMBH binaries, for the mass parameters ((a) + (b)), the distance (c), the sky localization (d), the inclination angle (e), and the polarization angle (f). In this plot, BD, BD2 stand for BDEC, BDEC2.}
   \label{fig:ViolinIMBH}
\end{figure*}

Now we take a look at the average parameter estimates for the cosmological distribution; the estimates are shown in tab.~\ref{table:avergageerrorsIMBH}. We see the same effects as in the previous section - due to the lower average SNR of the binaries in the cosmological distribution, some of the error estimates are orders of magnitude larger. In contrast to the example event, the sky localization, the distance, the inclination angle and the polarization angle in particular remain largely undetermined, with the notable exception of BDEC. The differences between BDEC and BDEC2 are much larger compared to the stellar-mass BBH case. LISA is able to determine the sky localization, but much less precisely than BDEC.

In fig.~\ref{fig:ViolinIMBH}, the distributions of the parameter estimates are shown. Here, in the case of IMBHs, we can see an effect in BDEC2 which is similar to what we saw in the case of stellar-mass black-hole binaries in ET. In BDEC2 the spread of the parameter estimates of the sky localization, the inclination and the polarization angles and the distance is rather wide. For example, there is little information about the distance in most events, but for some selected events, the estimates are very precise.

Summing up the results, we have seen that B-DECIGO is by far the detector best suited to observe IMBH binaries. It is the most sensitive detector with the most precise parameter estimates. ET is only able to detect lighter IMBH binaries, heavier binaries merge at frequencies below the ET sensitivity. With respect to multiband parameter estimation, we can improve the mass estimate significantly if we combine information from LISA with information from ET. Both B-DECIGO and LISA are able to determine the sky localization before the merger.

\section{Neutron Star binaries: LIGO/Virgo, ET, and B-DECIGO}
\label{sec:NSbinaries}

\subsection{Detection capabilities of the detectors}

\begin{figure}[t]
\centering
\includegraphics[scale=0.5]{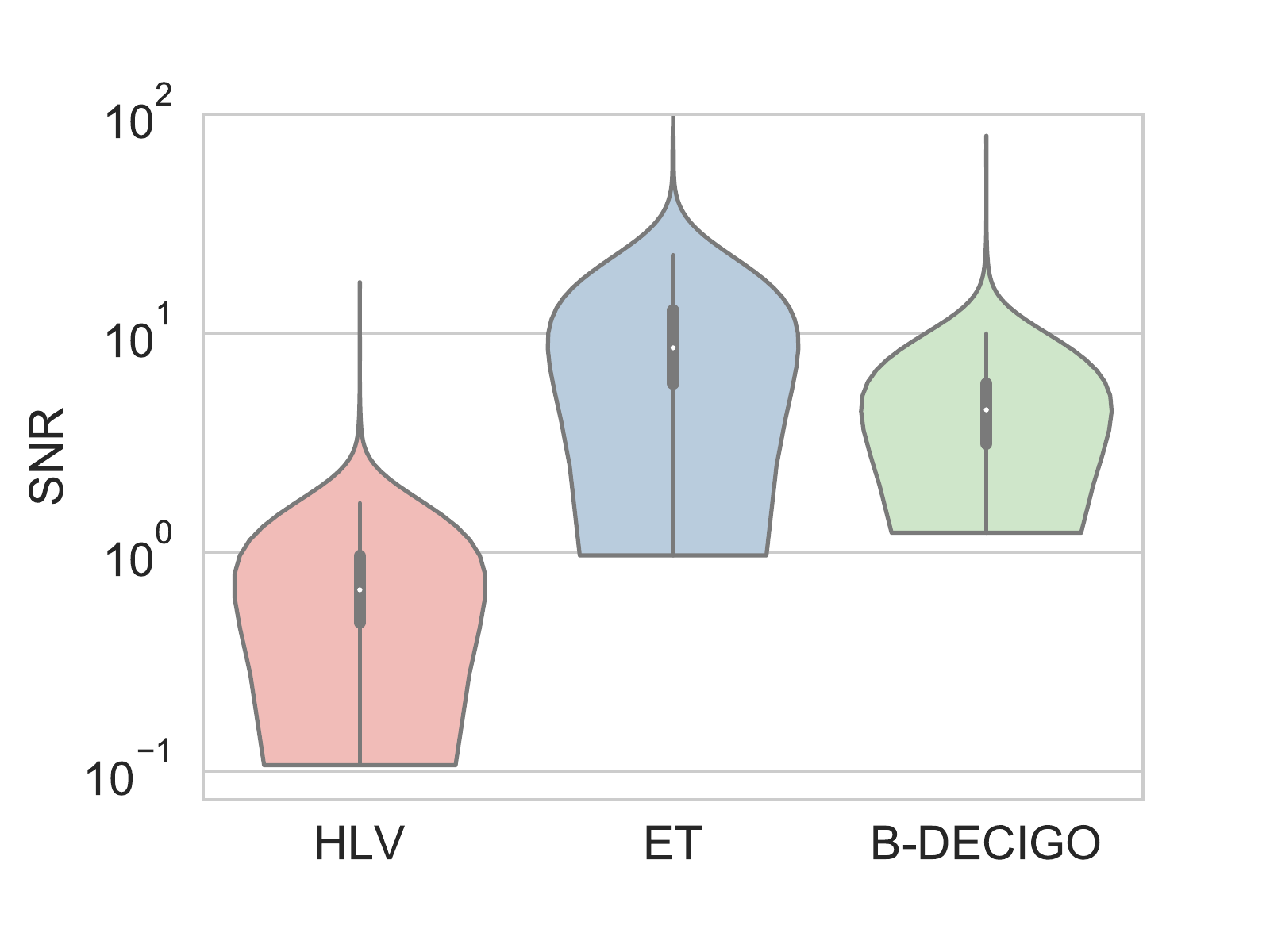}
\caption{Distribution of SNRs the LIGO/Virgo detectors, ET, and B-DECIGO, for a cosmological distribution of 1000 neutron star binaries with redshift $z<1$.}
\label{fig:SNRdistNSbinaries}
\end{figure}
\begin{table}
\begin{tabular}{c|c|c|c}
$\text{SNR}>10$& HLV & ET & B-DECIGO \\ 
\hline 
out of 1000 events & 1 & 378 & 46 \\ 
\end{tabular}
\caption{Number of neutron-star binary merger events with $\text{SNR}>10$ in different detectors.}
\label{table:eventsNSbinaries}
\end{table}

We begin by performing a population study of 1000 binary neutron stars (BNSs) to see what the detection capabilities of the detectors are. We assume that the mass spectrum of neutron stars in binaries is independent of redshift \cite{MaEA2019}. For the spatial distribution, we use the same model as before (see fig.~\ref{fig:Pz}), but we restrict the redshifts to $z<1$, because BNSs are only detected at small redshifts, due to lower SNRs. For the masses we choose a Gaussian distribution centered around $m = 1.33 \msolar$, with a standard deviation $\sigma_\text{m} = 0.09 \msolar$, and as minimum value we choose $m_\text{min} = 1 \msolar$, as maximum value $m_\text{max} = 2 \msolar$. The SNR distribution is shown in fig.~\ref{fig:SNRdistNSbinaries}. The number of events with $\text{SNR}>10$ are shown in table~\ref{table:eventsNSbinaries}. ET is the detector that is suited best for the detection of BNSs (378 out of 1000 events detected); B-DECIGO is less sensitive, but it still detects a number of events (46 out of 1000 events detected). The currently operating HLV detectors are, in comparison, not very sensitive, and detect only 1 out of 1000 events.

\subsection{An example event}

\begin{figure}[t]
\centering
\includegraphics[scale=0.5]{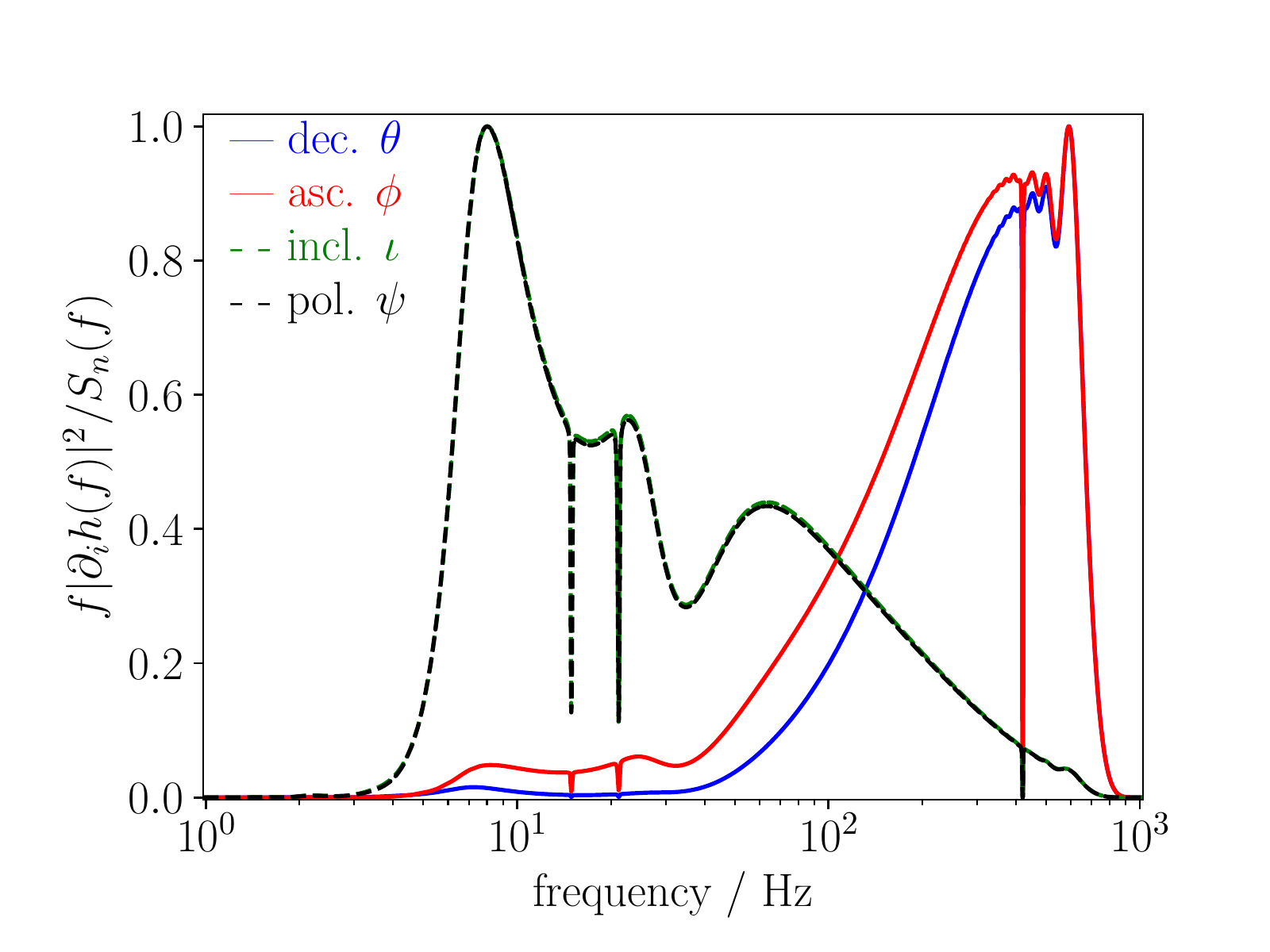}
\caption{Frequency dependence of the information content of the diagonal element of the Fisher matrix in ET. The spikes are due to ET's sensitivity curve, see fig.~\ref{fig:noisecurves}.}
\label{fig:NSETinfoplot}
\end{figure}

Now let us take a look at an example event. The event we are looking at is the merger of a BNS with a neutron star of mass $m_1 = 1.2 \msolar$ and another neutron star of the mass $m_2 = 1.4 \msolar$, at a distance of $r = 100 \Mpc$. The SNR of this event in the detectors is shown in table~\ref{table:SNRNSbinary}. The estimates for the standard deviations is shown in table~\ref{table:errorsNSbinary}.
\begin{table}
\begin{tabular}{c|c|c|c|c}
• & HLV & ET & BDEC & LISA \\ 
\hline 
SNR$^{\phantom{P^P}}$ & 19 & 333 & 143 & $4\times 10^{-2}$ \\ 
\end{tabular}
\caption{SNRs of the example binary in the detectors.}
\label{table:SNRNSbinary}
\end{table}
As we can see in table~\ref{table:SNRNSbinary}, the LIGO/Virgo detectors, ET and B-DECIGO can detect the binary with a sufficiently high SNR for a Fisher analysis. In LISA, however, the SNR is insufficient for a multiband observation. This was to be expected; already for stellar-mass BBHs LISA's sensitivity is too low to participate in multiband observations at least if only the last three years of these BBHs in the LISA band are used. Neutron star binaries are significantly lighter, and thus even harder to detect. This is not to say that LISA will not detect a neutron star binary. As summarized in \cite{KuEA2018}, we know of compact-binary systems, especially including white-dwarfs, that LISA will be able to detect. Therefore, it seems plausible to assume that a neutron-star binary exists in our galaxy or nearby galaxies (somewhat higher in frequency than, for example, J0737–3039A/B \cite{Lyne1153}), which can be detected in LISA. Nevertheless, this binary will not chirp to the frequencies of the ground-based detectors in a sufficiently short time, which makes it uninteresting for multiband parameter estimation.
\begin{table*}
\begin{tabular}{c|c|c|c|c|c|c|c|c}
• & HLV & ET & BDEC  & BDEC2 & HLV/BDEC & ET/BDEC & HLV/ET  & HLV/ET/BDEC \\ 
\hline 
$\Delta \mu$ / $\msolar$ & $1 \times 10^{-3}$  & $2 \times 10^{-5}$ & $1.3 \times 10^{-5}$ & $1.4 \times 10^{-5}$ & $1.2 \times 10^{-5}$  & $6.2 \times 10^{-6}$ & $2 \times 10^{-5}$ & $6.2 \times 10^{-6}$ \\ 
\hline 
$\Delta \Mtot$ / $\msolar$ & $6.1 \times 10^{-3}$  & $1.2 \times 10^{-4}$ & $7.6 \times 10^{-5}$ & $8.4 \times 10^{-5}$ & $7 \times 10^{-5}$  & $3.7 \times 10^{-5}$ & $1.2 \times 10^{-4}$ & $3.7 \times 10^{-5}$ \\ 
\hline 
$\Delta r$ / $\Mpc$ & $68$ & $2.48$ & $3.71$ & $4.3$ & $3.69$ & $1.38$ & $2$ & $1.38$ \\ 
\hline 
$\Delta \iota / \deg$ & $42$ & $2.28$ & $2.43$ & $2.82$ & $2.41$ & $0.96$ & $1.95$ & $0.96$ \\ 
\hline 
$\Delta \theta / \deg$ & $1.44$ & $1.13$ & $1.7 \times 10^{-3}$ & $1.9 \times 10^{-3}$ & $1.7 \times 10^{-3}$ & $1.3 \times 10^{-3}$ & $0.64$ & $1.3 \times 10^{-3}$ \\ 
\hline 
$\Delta \phi / \deg$ & $1.31$ & $1.06$ & $1 \times 10^{-3}$ & $1.2 \times 10^{-3}$ & $3.6 \times 10^{-4}$ & $2.2 \times 10^{-4}$ & $0.56$ & $2.2 \times 10^{-4}$  \\ 
\hline 
$\Delta \psi / \deg$ & $81$ & $3.6$ & $3.4$ & $3.6$ & $3.4$ & $1.4$ & $1.7$ & $1.4$ \\
\hline 
$\Omega / \deg^2$ & $2.75$ & $3.75$ & $5.5 \times 10^{-6}$ & $7 \times 10^{-6}$ & $1.8 \times 10^{-6}$ & $8.9 \times 10^{-7}$ & $0.5$ & $8.9 \times 10^{-7}$ \\
\end{tabular}
\caption{Fisher estimates for the standard deviations of the neutron star example binary, for different detectors. BDEC1 is B-DECIGO with a geostationary orbit, BDEC2 with a LISA-like orbit.}
\label{table:errorsNSbinary}
\end{table*}
Taking a look at the standard deviations in table~\ref{table:errorsNSbinary}, it is interesting to observe the effect of the long observation time of BNSs. When observing BNSs, ET is able to determine the sky localization, because the binary spends more than a day in the frequency range of ET. The detector motion breaks the parameter degeneracy. We again plot $f|\partial_i h(f)|^2/S_n(f)$ for the position-, inclination-, and polarization angles, in fig.~\ref{fig:NSETinfoplot}. The sky localization in ET is sensitive to the mass of the BNS, because lighter BNSs spend more time emitting in the ET frequency range, which means that the detector movement effects are more important.

When comparing BDEC and BDEC2, we see that for BNSs, the differences between geostationary and LISA-like orbit become very small. BDEC is still slightly better than BDEC2, but the effect is small. This is what we expected: BNSs spend several years in the B-DECIGO frequency range, which is longer than the time scale of the detector motion of BDEC2. Therefore, the detector motion of BDEC2 breaks the parameter degeneracy almost as well as BDEC. In the example binary, apart from the sky localization, ET and BDEC give comparable parameter estimates. The sky localization estimate is much better in BDEC. HLV gives a sky localization estimate comparable to the one provided by ET, for all other parameters the estimate provided by the HLV detectors is much less precise.

When considering the detector networks, we see again that the mass estimates benefit from combining space-borne with ground-based detectors. With respect to the sky localization, we see that there is a significant improvement when combining the HLV detectors with ET (timing triangulation with four detectors). For most parameters, the estimates of the detectors networks are generally close to the estimate of the best detector in the network, and improve by a factor of a few at most.

It is very promising to see that B-DECIGO is able to obtain information about the sky localization before the merger takes place, which will allow for multi-messenger astronomy. When the BNS leaves the B-DECIGO frequency band, a few minutes remain until the merger takes place, but the sky localization can be given earlier. For particularly close and bright events, this may even be possible with ET. This is particularly important, because BNSs are the sources that are expected to have an electromagnetic counterpart.

\subsection{Average parameter estimates}

\begin{table*}
\begin{tabular}{c|c|c|c|c|c|c|c|c}
• & HLV & ET & BDEC  & BDEC2 & HLV/BDEC & ET/BDEC & HLV/ET  & HLV/ET/BDEC \\ 
\hline 
$\Delta \mu$ / $\mu$ & $2.3 \times 10^{-3}$  & $7.5 \times 10^{-5}$ & $4.4 \times 10^{-5}$ & $4.5 \times 10^{-5}$ & $3.9 \times 10^{-5}$  & $2.3 \times 10^{-5}$ & $7.3 \times 10^{-5}$ & $2.3 \times 10^{-5}$ \\ 
\hline 
$\Delta \Mtot$ / $\Mtot$ & $3.3 \times 10^{-3}$  & $1.1 \times 10^{-4}$ & $6.7 \times 10^{-5}$ & $6.7 \times 10^{-5}$ & $5.9 \times 10^{-5}$  & $3.4 \times 10^{-5}$ & $1.1 \times 10^{-4}$ & $3.4 \times 10^{-5}$ \\ 
\hline 
$\Delta r$ / $r$ & $1.44$ & $0.48$ & $0.49$ & $0.53$ & $0.47$ & $0.33$ & $0.41$ & $0.33$ \\ 
\hline 
$\Delta \cos(\iota)$ & $1.39$ & $0.46$ & $0.47$ & $0.52$ & $0.46$ & $0.32$ & $0.4$ & $0.33$ \\ 
\hline 
$\Delta \theta / \deg$ & $2.3$ & $3$ & $0.013$ & $0.014$ & $6.6 \times 10^{-3}$ & $4.7 \times 10^{-3}$ & $0.81$ & $4.8 \times 10^{-3}$ \\ 
\hline 
$\Delta \phi / \deg$ & $1.8$ & $3.5$ & $3 \times 10^{-3}$ & $3.8 \times 10^{-3}$ & $1.8 \times 10^{-3}$ & $1.6 \times 10^{-3}$ & $0.84$ & $1.6 \times 10^{-3}$ \\ 
\hline 
$\Delta \psi / \deg$ & $240$ & $122$ & $117$ & $122$ & $116$ & $84$ & $95$ & $83$ \\
\hline 
$\Omega / \deg^2$ & $10$ & $47$ & $4.3 \times 10^{-5}$ & $6 \times 10^{-5}$ & $1.2 \times 10^{-5}$ & $6.9 \times 10^{-6}$ & $1.6$ & $6.9 \times 10^{-6}$ \\
\end{tabular}
\caption{Averaged Fisher estimates for the 26 BNSs with SNR $>10$ in all detectors or detector networks included in this table. Here, the 26 signals resulted from a larger cosmological distribution of 100000 binaries to have a sufficiently high number of detected signals to average over.}
\label{table:averageerrorsNSbinary}
\end{table*}

\begin{figure*}[!ht]
   \centering
   (a)\includegraphics[width=.45\textwidth]{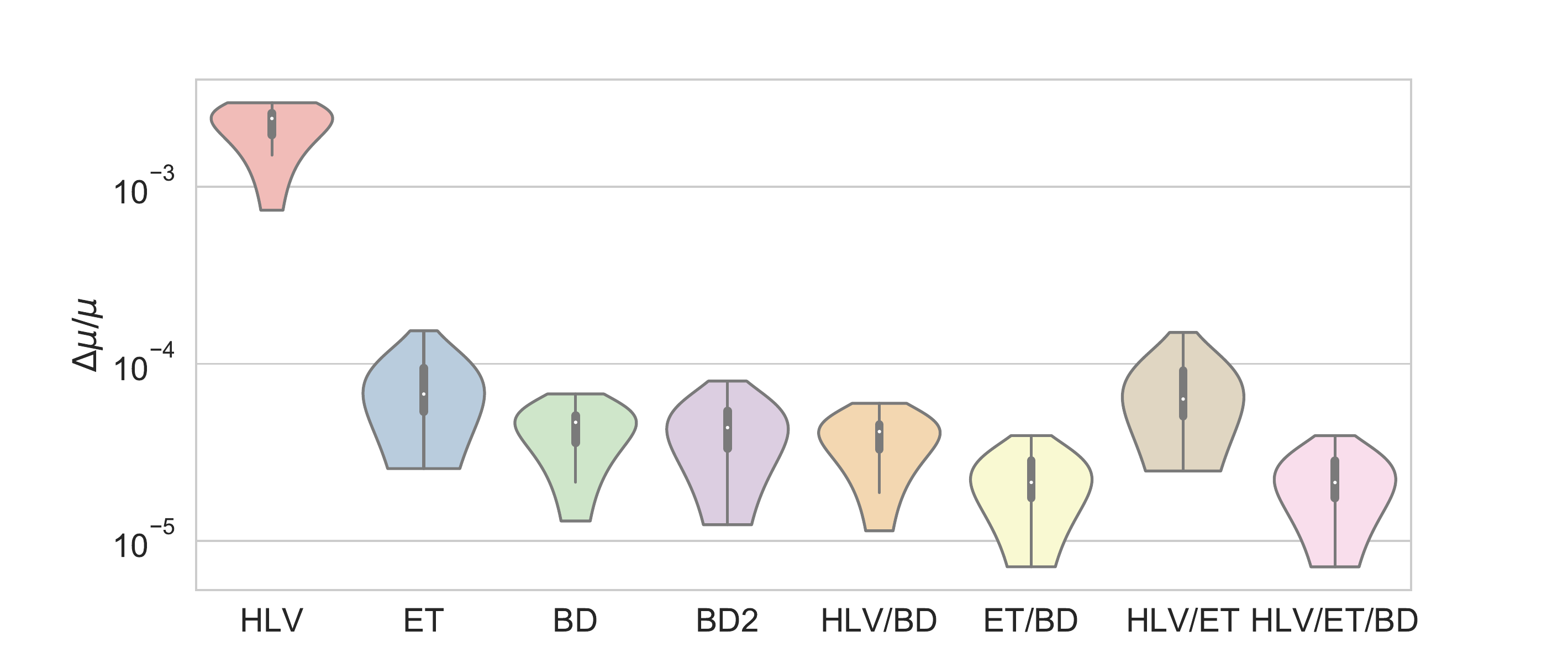}
   (b)\includegraphics[width=.45\textwidth]{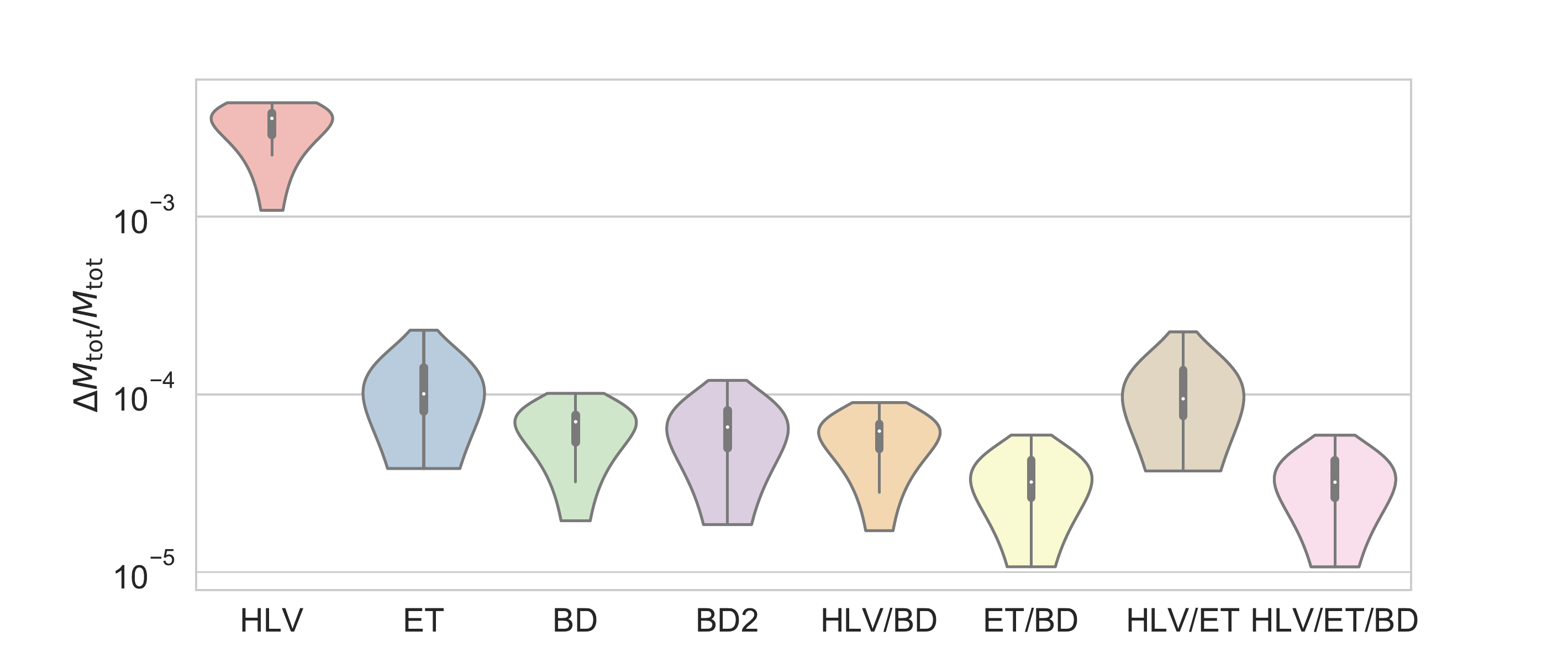}\\
   (c)\includegraphics[width=.45\textwidth]{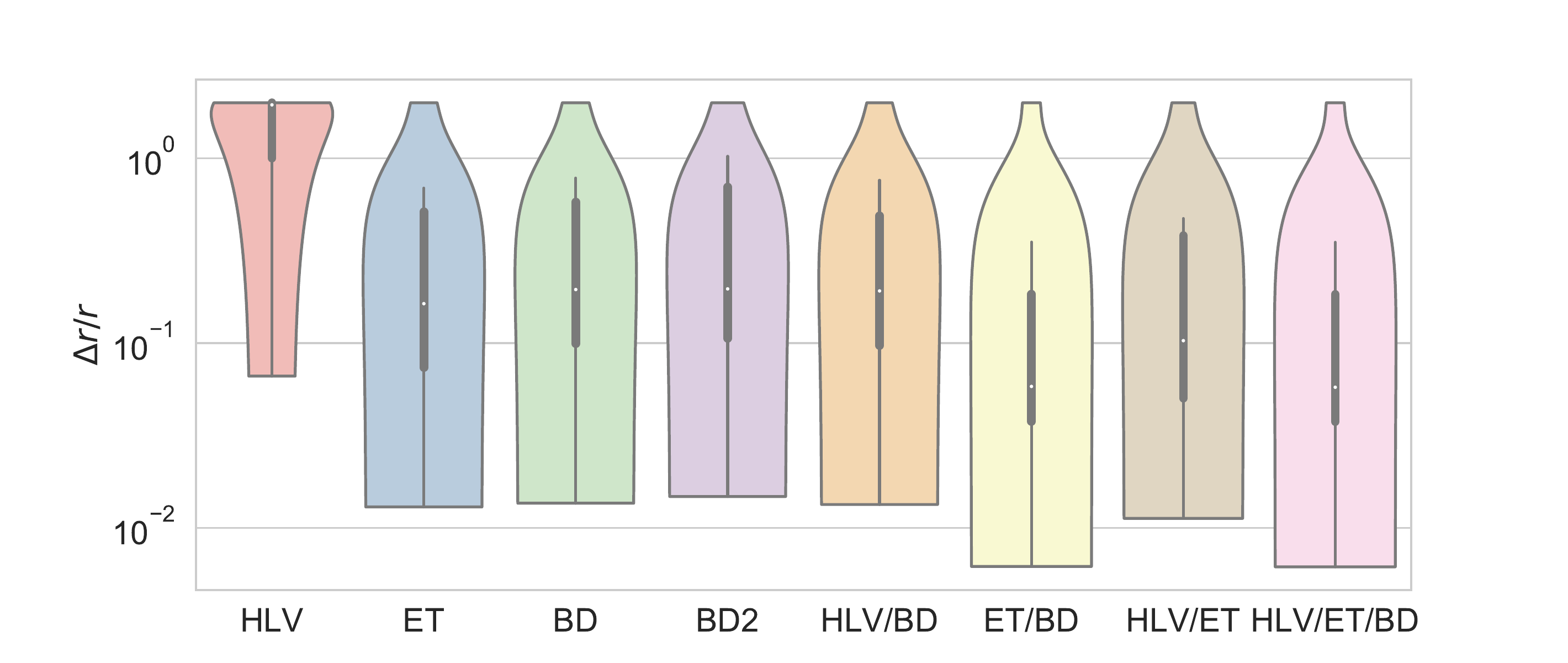}
   (d)\includegraphics[width=.45\textwidth]{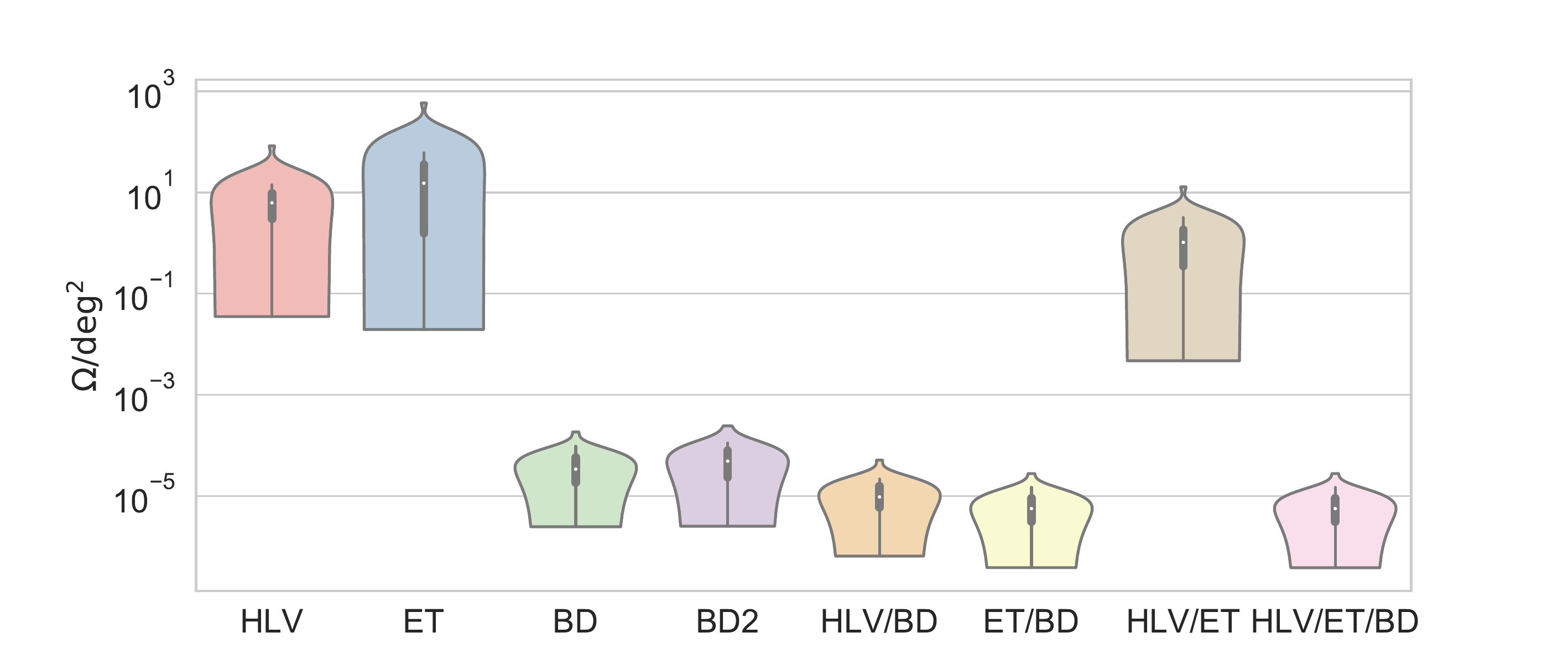}\\
   (e)\includegraphics[width=.45\textwidth]{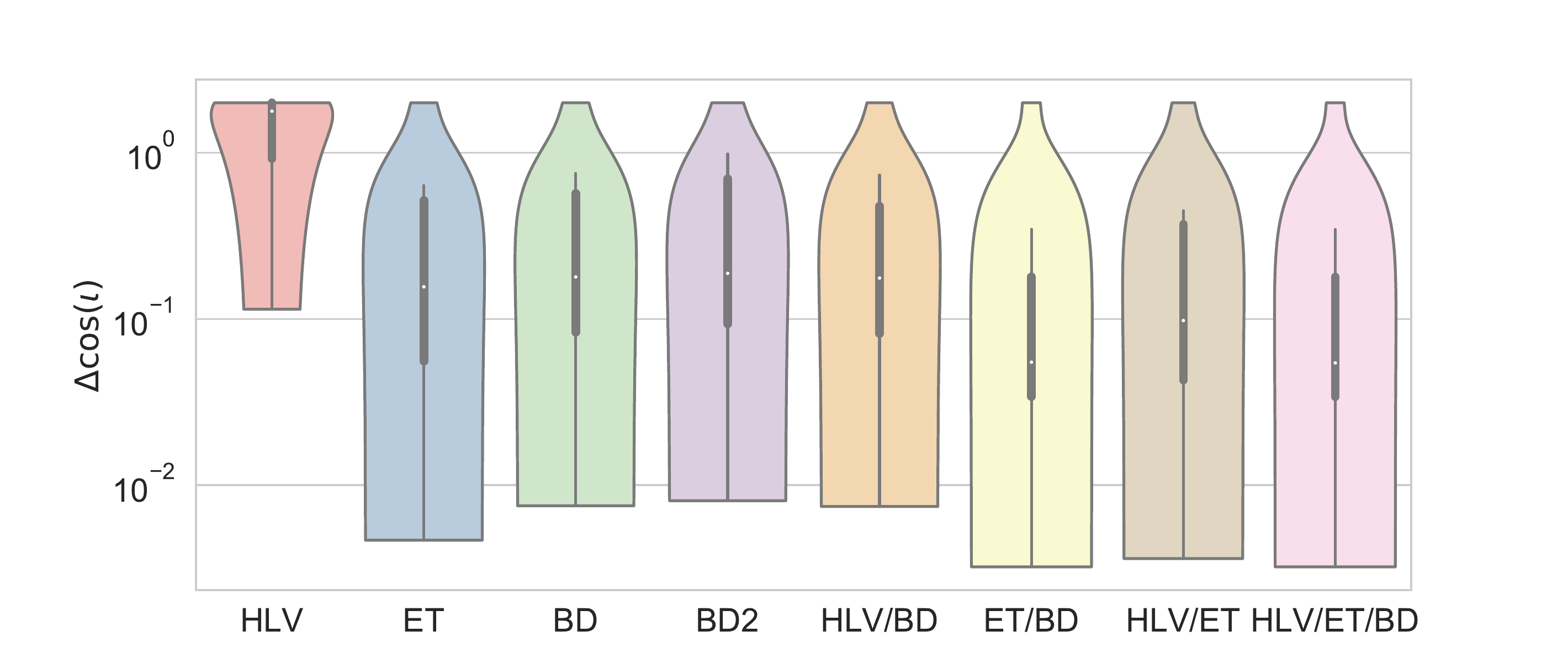}
   (f)\includegraphics[width=.45\textwidth]{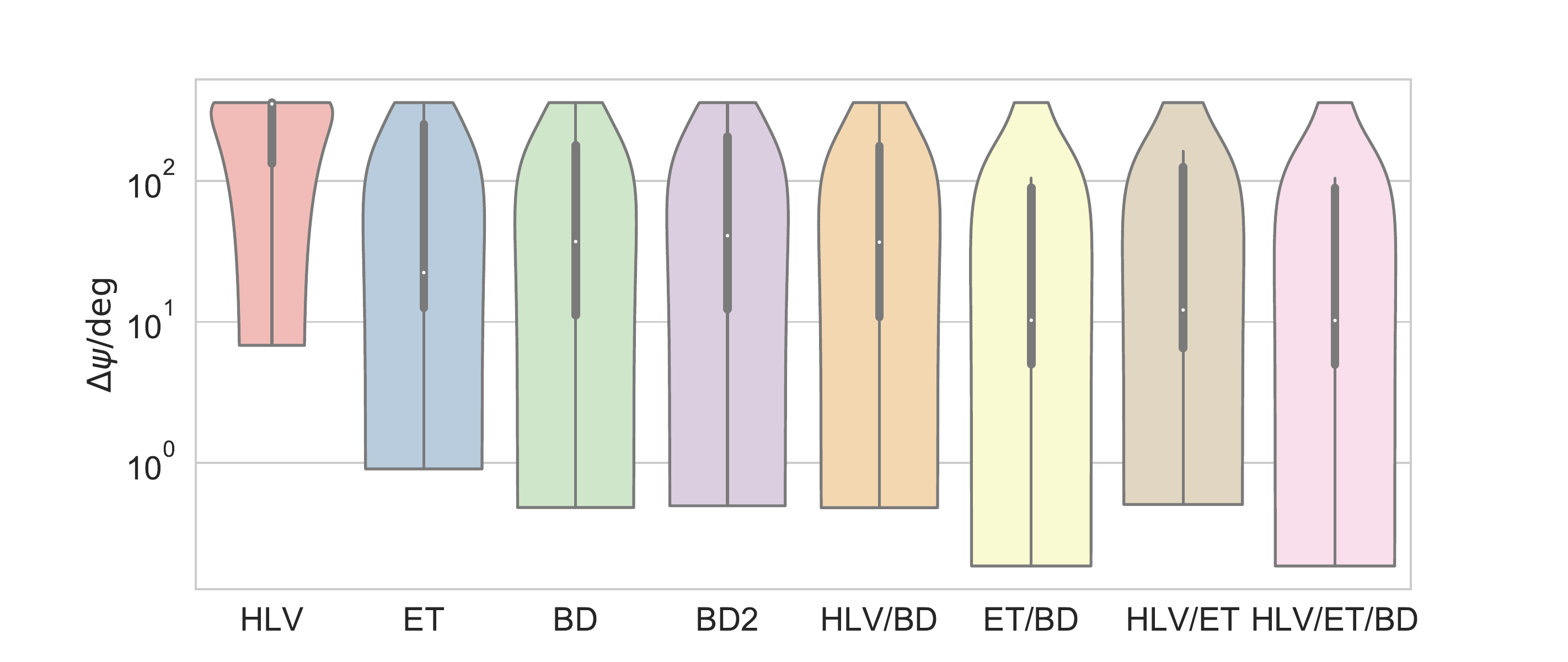}\\
   \caption{Distributions of the standard deviations of the parameter estimate of the neutron-star binaries, for the mass parameters ((a) + (b)), the distance (c), the sky localization (d), the inclination angle (e), and the polarization angle (f). In this plot, BD, BD2 stand for BDEC, BDEC2.}
  \label{fig:ViolinNS}
\end{figure*}

Next, we take a look at the average standard deviations. Again, we averaged the estimates of a cosmological distribution of BNSs. The results are shown in table~\ref{table:averageerrorsNSbinary}. Since our initial cosmological distribution with 1000 sources only led to one BNS detection with HLV, we here consider a larger distribution with 100000 signals. This leads to 26 signals detected with HLV (and therefore detected in ET and B-DECIGO as well). In the HLV detectors, B-DECIGO and ET, the average sky localization is good enough for multi-messenger studies, with B-DECIGO giving the best estimate of the sky localization. We can also see that in all detectors, the average standard deviations of the inclination angle, the polarization angle and the distance are rather large. 

In fig.~\ref{fig:ViolinNS}, we show the distributions of the parameter estimates. We see that the sky localization estimates in ET vary greatly --- many events have a good sky localization of around $10\,\deg^2$--$100\,\deg^2$, but there are also some events with sky localizations of a few $100\,\deg^2$, i.e., too large to allow for multi-messenger studies. The estimates of the inclination- and polarization angle and the distance vary greatly as well. It seems that the estimate of this parameter depends strongly on the specifics of the individual binary.

\subsection{ET: Pre-merger sky localization information}

Since ET is able to detect a large number of BNSs, and since the previous section suggests that it might be able to gain sky localization information before the merger takes place, a further investigation of this is warranted. To this end, we simulate the waveform only up to 8 minutes before merger. This leaves some time for data analysis and for sending a warning to the telescopes. The results of the parameter estimation of the example binary considered before are shown in table~\ref{table:ETpremerger}.
\begin{table}
\begin{tabular}{c|c|c|c|c|c}
Pre-merger & SNR & $\Delta r / \Mpc$ & $\Delta \theta / \deg$ & $\Delta \phi / \deg$ & $\Omega / \deg^2$ \\ 
\hline 
estimate & 206 & 3.6 & 1.35 & 1.31 & 5.5 \\ 
\end{tabular}
\caption{Pre-merger estimates with ET for the example binary.}
\label{table:ETpremerger}
\end{table}
\begin{figure}[t]
\centering
\includegraphics[scale=0.5]{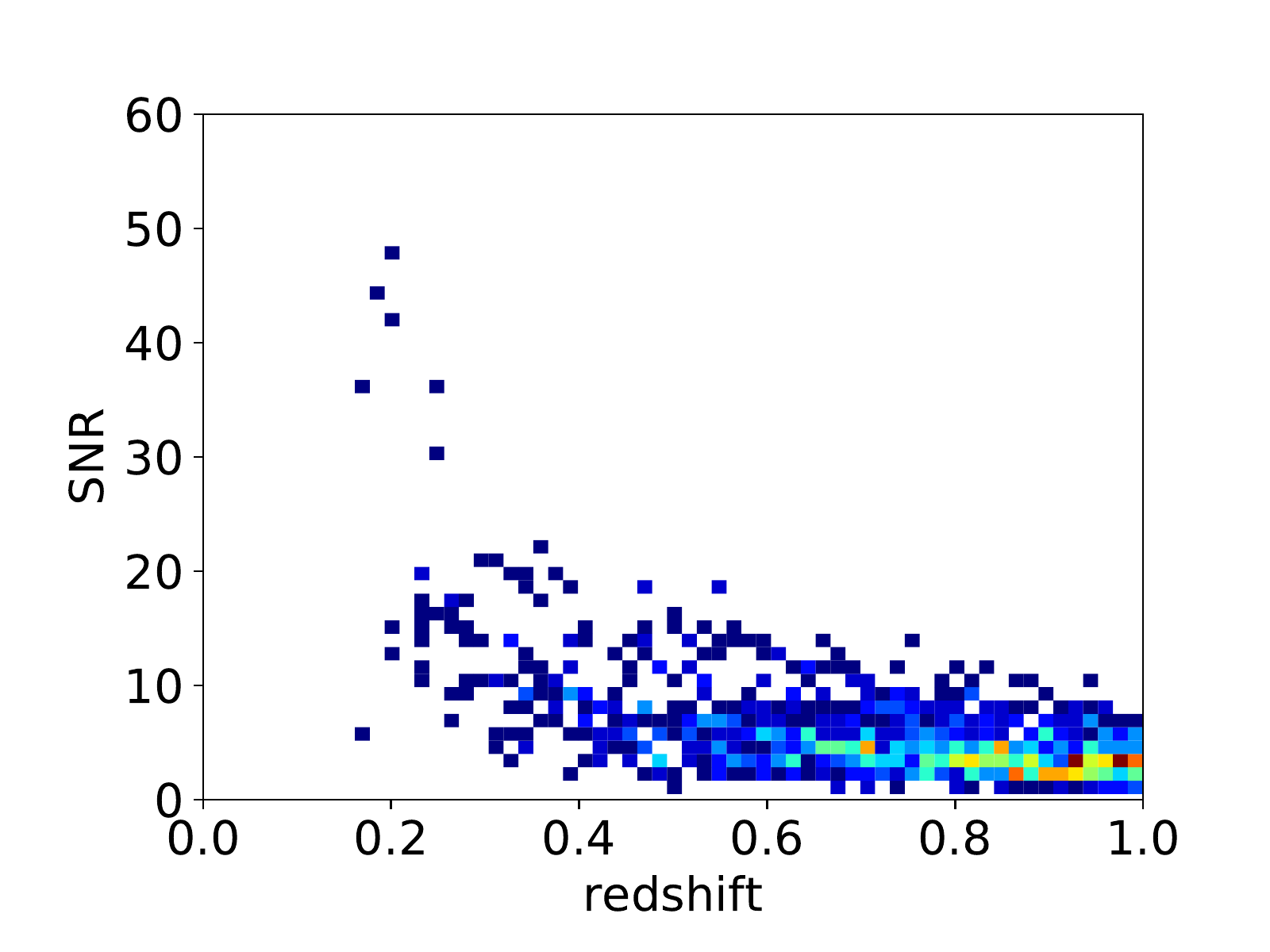}
\caption{ET SNR distribution of a cosmological distribution of 1000 neutron-star binaries with redshift $z<1$, where the SNR is integrated until 8 minutes before the merger (premerger condition).}
\label{fig:SNRdistNeutronStarETpremerger}
\end{figure}

\begin{figure}[t]
\centering
\includegraphics[scale=0.5]{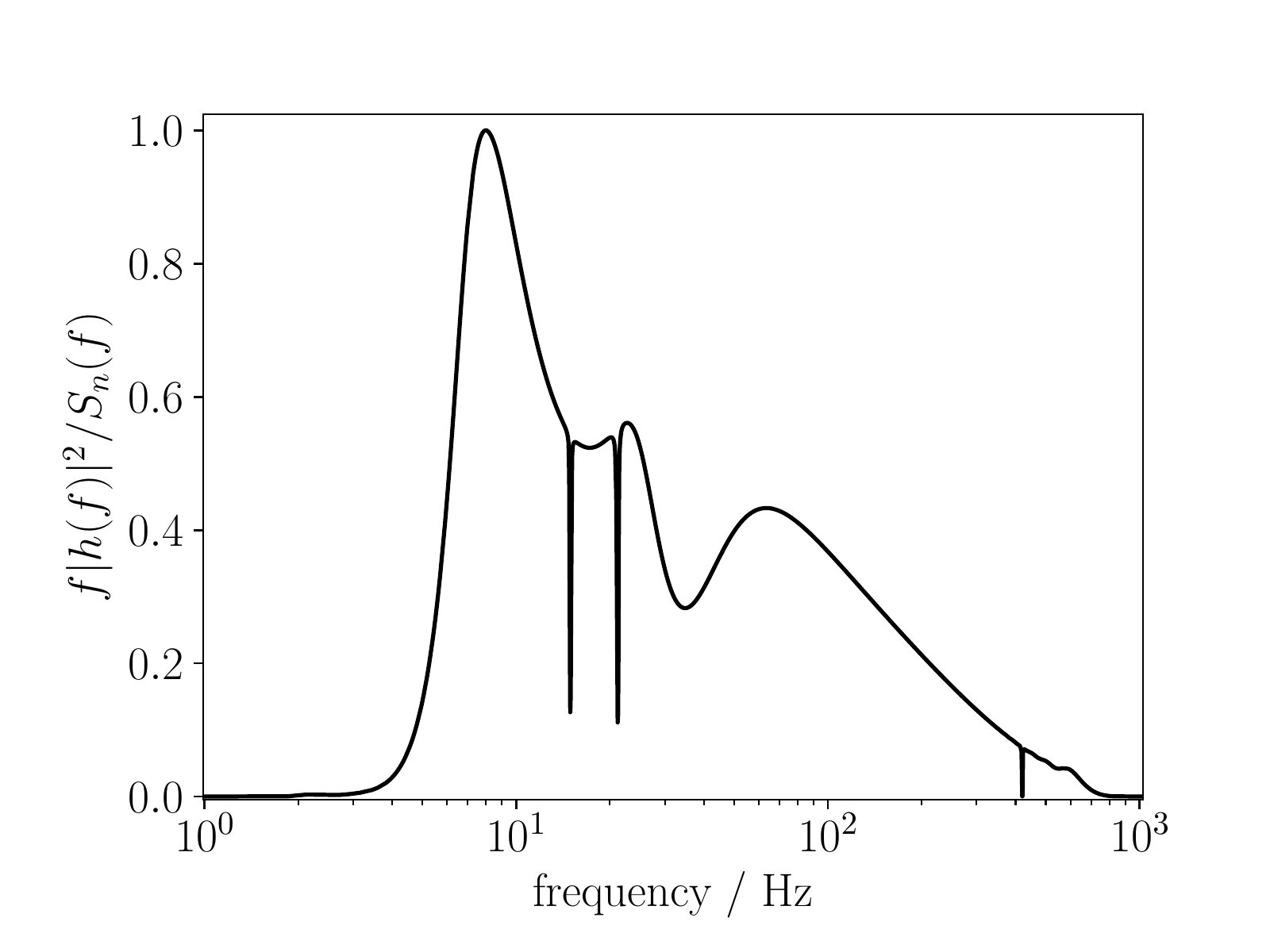}
\caption{Contribution to the SNR in ET from different frequencies, in the example BNS.}
\label{fig:SNRET}
\end{figure}

\begin{table}
\begin{tabular}{c|c|c|c|c}
Pre-merger & $\Delta r / r$ & $\Delta \theta / \deg$ & $\Delta \phi / \deg$ & $\Omega / \deg^2$ \\ 
\hline 
$\text{SNR}>10$ & 1.25 & 100 & 107 & 19000 \\ 
\hline 
$\text{SNR}>20$ & 0.69 & 40 & 59 & 8000 \\ 
\hline 
$\text{SNR}>40$ & 0.88 & 15 & 53& 5500 \\ 
\end{tabular}
\caption{Average error estimates for the 8\,min pre-merger sky localization with ET, for events with $\text{SNR}>10$ (114 events), $\text{SNR}>20$ (13 events), and $\text{SNR}>40$ (5 events).}
\label{table:ETpremergeraverage}
\end{table}
We can see that, compared to full-signal ET considered in the previous section, the SNR is reduced, but still significantly higher than the SNR of the HLV detectors. Due to the reduced SNR, the sky localization estimate is also worse than in the full-signal case, but it is still good enough for multimessenger astronomy.

Figure \ref{fig:SNRdistNeutronStarETpremerger} shows the SNR distribution in ET under pre-merger alert conditions for a cosmological distribution of 1000 BNSs; 114 out of 1000 have an SNR larger than 10. Table \ref{table:ETpremergeraverage} shows the estimates averaged over the binaries from the cosmological distribution, for different SNR thresholds. If we impose the condition $\text{SNR}>10$, we have 114 events, for $\text{SNR}>20$ we have 13 events, and for $\text{SNR}>40$ we have 5 events. We can see that we do obtain some information on the sky localization, but it is not sufficient for pointing telescopes into the right direction. Increasing the SNR threshold selects the closer and brighter events, but even for $\text{SNR}>40$ the sky localization is not very precise. Only with exceptionally bright and close binaries like the example binary we can hope to constrain the sky localization before merger with ET sufficiently to allow for multi-messenger studies. This is in harmony with the findings of \cite{earlywarning}, which provides a more detailed study.

Figure~\ref{fig:SNRET} shows the contribution to the SNR from different frequencies, in the case of the example binary. The exact form depends on the detector orientation over the course of the observation time. If we impose the premerger condition of a warning time of 8 minutes, we loose the SNR from the frequencies above the gravitational wave frequency emitted 8 minutes before merger, and this frequency is typically in the range of 5 -- 20\,Hz, depending on source distance and intrinsic mass. It is also clear that the SNR reduction depends very strongly on the warning time. Reducing the warning time can significantly improve the SNR as well as the sky localization estimate.

\section{Conclusions}

In this study we have investigated the abilities of future detectors to determine parameters of gravitational wave sources. We have seen that the time scales of the detector motion and the chirp of the source are crucial factors. The motion of the detectors breaks the degeneracies of the parameters and vastly improves the parameter estimates, in particular of the source position-, the inclination- and the polarization angles. The condition for this effect to kick in is that the time scale of the detector motion is smaller than the time in which the source emits gravitational waves in the frequency range of the detector. Since the merger frequencies and the speed of the chirp depend on the mass of the astrophysical object, the ability of a detector to reconstruct the source parameters strongly depends on the mass of the source. Regarding the different possible orbits of the B-DECIGO detector, we have seen that B-DECIGO with a geostationary orbit provides better estimates of the sky localization and the inclination- and polarization angles, due to the shorter timescale of the detector motion. B-DECIGO would also be able to detect almost all the IMBHs in the universe, and would therefore answer the question whether they exist or not. With respect to LISA, we have seen that its sensitivity is not high enough to detect a sufficient number of stellar-mass BBHs - we do not expect LISA to be a good candidate detector for multiband parameter estimation of sources of this kind. IMBHs, if they do exist, are more promising sources for LISA, since their signals are much stronger, due to the higher masses.

With respect to ET, we have seen that the triangular shape gives rise to interesting effects when estimating the sky localization of stellar-mass black hole binaries. While ET is unable to reconstruct the sky localization of the vast majority of black hole binaries, for a few selected ones it is able to determine the sky localization well. It seems plausible that this is due to the triangular shape of the detector, which makes it possible to determine the polarization and break some degeneracies, even without long observation times or time triangulation with a network.
Due to the long observation time and the detector movement, ET can estimate the sky localization of neutron star binaries.
We have also investigated whether ET can constrain the sky localization of neutron star binaries before the merger takes place, which would be useful for the purpose of multi-messenger studies. We have seen that this is possible only for very close neutron star binaries.

The estimate of the mass parameters benefits greatly from combining space-borne detectors with ground-based detectors; this is because the estimate of the mass parameters in space-borne detectors is limited by their degeneracy in the early inspiral. Additional information about the merger waveform from ground-based detectors can improve the estimate significantly.

\section*{Acknowledgements}

The authors would like to thank Michela Mapelli for her guidance on the sparse information available about intermediate-mass Binary Black Holes. We also want to express our gratitude to Seiji Kawamura, Naoki Seto, and Sato Shuichi for taking the time to explain the current status of the DECIGO detector, as well as providing information about the sensitivity curves and possible orbits.

\bibliographystyle{apsrev} 
\bibliography{references}

\end{document}